\documentclass[a4paper,12pt]{article}
\pdfoutput=1 

\usepackage{a4wide}
\usepackage{cite}

\usepackage[T1]{fontenc} 
\usepackage[latin1]{inputenc}

\usepackage{graphicx}
\usepackage{epstopdf} 

\usepackage{empheq}
\usepackage{amssymb}
\usepackage{amsmath}
\usepackage{amsfonts}
\usepackage{xcolor}

\DeclareMathAlphabet{\mathpzc}{OT1}{pzc}{m}{it}

\numberwithin{equation}{section}

\usepackage[colorlinks]{hyperref}

\hypersetup{
 citecolor=blue,
 linkcolor=blue,
 urlcolor=blue}

\def \pb {{\rm pb}}

\def \wt {{\widetilde w}}
\def \Li {{\rm Li}}

\def \FLM {F_{\rm LM}}
\def \FLMns {F_{\rm LM,ns}}
\def \FLMs {F_{\rm LM,s}}

\def \FLMg {F_{{\rm LM},g}}
\def \FLV {F_{\rm LV}}
\def \FLVg {F_{{\rm LV},g}}
\def \FLVf {F_{\rm LV}^{\rm fin}}
\def \FLVfg {F_{{\rm LV},g}^{\rm fin}}
\def \FLVVf {F_{\rm LVV}^{\rm fin}}
\def \FLVsqf {F_{\rm LV^2}^{\rm fin}}
\def \FLVV {F_{\rm LVV}}

\def \NSS {NSS}

\def \d {{\rm d}}
\def \dsh {{\d\hat\sigma}}
\def \LO {{\rm LO}}
\def \NLO {{\rm NLO}}
\def \NNLO {{\rm NNLO}}

\def \Cf{{C_F}}
\def \Ca{{C_A}}
\def \tr{{T_R}}
\def \nf {{n_f}}

\def \asontwopimu {{\frac{\alpha_s(\mu)}{2\pi}}}

\def \qas {{[\alpha_s]}}
\def \SS {S{\hspace{-5pt}}S}
\def \CC {C{\hspace{-6pt}C}}
\def \I {I}
\def \ONLO {\hat{\mathcal O}_{\rm NLO}}

\def \Em {E_{\rm max}}
\def \zm {z_{\rm min}}

\def \qb {{\bar q}}
\def \QB {{\bar Q}}

\def \PAP {{\hat P}}
\def \calP {{\mathcal P}}

\definecolor{darkblue}{rgb}{0,0,0.7}

\newcommand{\la}{\left\langle}
\newcommand{\ra}{\right\rangle}

\newcommand{\lp}{\left(}
\newcommand{\rp}{\right)}

\newcommand{\be}{\begin{equation}}
\newcommand{\ee}{\end{equation}}
\newcommand{\bes}{\begin{equation}\begin{split}}

\newcommand{\df}[1]{[df_{#1}]}

\newcommand{\ep}{\epsilon}

\newcommand{\DD}[1]{\mathcal D_{#1}}

\begin{document}
\vspace{-5.0cm}
\begin{flushright}
OUTP-19-12P, 
CERN-TH-2019-180, TTP19-035, P3H-19-040
\end{flushright}

\vspace{2.0cm}

\begin{center}
{\large \bf Analytic results for deep-inelastic scattering at NNLO QCD
  with the nested soft-collinear subtraction scheme}\\
\end{center}

\vspace{0.5cm}

\begin{center}
Konstantin Asteriadis$^{1}$, Fabrizio Caola$^{2}$, 
Kirill Melnikov$^{1}$, Raoul R\"ontsch$^{3}$.\\
\vspace{.3cm}
{\it
  {}$^1$Institute for Theoretical Particle Physics, KIT, Karlsruhe, Germany\\
  {}$^2$Rudolf Peierls Centre for Theoretical Physics, Clarendon Laboratory, Parks Road, Oxford OX1 3PU, UK \& 
Wadham College, Oxford OX1 3PN\\
{}$^3$Theoretical Physics Department, CERN, 1211 Geneva 23, Switzerland\\
}

\vspace{1.3cm}

{\bf \large Abstract}
\end{center}
We present analytic results that describe fully-differential NNLO QCD
corrections to deep-inelastic scattering processes within the nested
soft-collinear subtraction scheme.  This is the last building block
required for the application of this scheme to computations of NNLO QCD
corrections to arbitrary processes at hadron colliders.
\thispagestyle{empty}

\clearpage
{\hypersetup{linkcolor=black}
\tableofcontents
}
\thispagestyle{empty}
\clearpage

\pagenumbering{arabic}

\allowdisplaybreaks

\section{Introduction}
\label{sec:introduction}

In this paper, we apply the nested soft-collinear subtraction scheme
for NNLO QCD computations, introduced by some of us
in Ref.~\cite{Caola:2017dug}, to the deep-inelastic scattering (DIS) process
 $Pe \to e +X$.  We note right away that our goal is not DIS
phenomenology; rather, we would like to extend this subtraction
scheme to processes that involve QCD partons in both the
initial and the final state.

Compared to the cases of color singlet production~\cite{Caola:2019nzf}
and decay~\cite{Caola:2019pfz} that we studied earlier, the DIS process
requires us to deal with the situation where QCD partons in the
(leading order) hard processes are not back-to-back.  This makes the
computation of NNLO QCD corrections to deep-inelastic scattering an important
next step in the development of the nested subtraction scheme (\NSS).

In spite of the fact that the computation that we report in this paper
is new, we would like to emphasize that we can re-use significant
parts of the analytic computations described in
Refs.~\cite{Caola:2019nzf,Caola:2019pfz}. This is so because collinear
singularities in QCD factorize on external lines so that their
treatment, including analytic integration of respective subtraction
terms \cite{maxtc}, is process-independent.  Hence, everything that
needs to be known about collinear singularities in DIS and their
regularization can be inferred from the treatment of the collinear
singularities in color-singlet production and color-singlet decays,
see Refs.~\cite{Caola:2019nzf,Caola:2019pfz}.

At variance with collinear limits, important differences arise in the
treatment of the (double-) soft radiation which is sensitive to the
relative orientation of three-momenta of hard emittors.  The
integrated double-soft subtraction term for the case when the momenta
of hard emittors are at an angle to each other was analytically
computed in Ref.~\cite{max_soft}.  The computation of NNLO QCD
corrections to the DIS process that we report in this paper is the first
application of that result.

The main result of this paper is the set of analytic formulas that,
in conjunction with the fully-resolved regulated contribution,
provides a fully-differential description of DIS at NNLO QCD.  It is
our long-term goal to employ these formulas as
ingredients to describe ``initial-final dipole'' contributions when
computing NNLO QCD corrections to generic processes. Because of that,
it is important to ensure that the analytic results for
initial-final dipoles reported in this paper are correct.  Studies
of DIS are advantageous from this perspective since analytic results
for DIS coefficient functions at NNLO are
available~\cite{Kazakov:1990fu,Zijlstra:1992qd,Moch:1999eb} and we can
use them to check our computations to a very high precision.

We note in passing that, in the past decade, a large number of
subtraction schemes and slicing methods appeared \cite{ant, czakonsub,
  czakonsub4d,Boughezal:2011jf, Cacciari:2015jma, qt1, qt2, njet1,
  njet2,Campbell:2017hsw,colorful,tackmann:nj:pow,frank:nj:pow,Ebert:2018gsn, magnea,
  herzog}; they enabled a large number
of NNLO QCD computations for important LHC processes
\cite{qt_diboson,qt_hh,qt_tt,mcfm_nnlo,njettiness_processes,czakon_top,czakon_jet,
antenna_hj,antenna_vj,antenna_jj,singletop,vbf,qt_vh,other_vh,us_hj}. Nevertheless, in
spite of all successes, the construction of a fully local, analytic,
physically transparent and scalable subtraction scheme remains an
interesting challenge. We believe that further development of the
{\NSS}, that we describe in this paper, will contribute to finding an
answer to this challenge.

The remainder of the paper is organized as follows. In
Section~\ref{sec:LOandNLO} we describe how leading order (LO) and
next-to-leading order (NLO) DIS cross sections are computed. In
Section~\ref{sec:NNLO} we discuss the NNLO computation.  In
Section~\ref{sec:results} we validate our results against analytic
ones.  We conclude in Section~\ref{sec:conclusion}. Useful formulas
are collected in several Appendices. Analytic results for NNLO
QCD DIS subtraction terms in computer-readable format are provided in
an ancillary file attached to this submission. 

\section{LO and NLO calculation}
\label{sec:LOandNLO}
We consider deep-inelastic scattering of an electron on a proton 
\be
P(P_1) + e(p_2) \to e(p_3) +  X,
\ee 
mediated by a neutral current. 
The cross section of this
process is computed as a convolution of parton distribution functions
with the partonic cross section that
describes parton-electron scattering. Schematically, we write 
\be {\rm d}
\sigma_H = \sum \limits_{i}^{} \;\int \limits_{0}^{1} {\rm d} x\;
f_i(x) \; {\rm d} \hat\sigma_{f_i+e \to X}(x P_1, p_2).
\label{eq2.2}
\ee 
In Eq.~(\ref{eq2.2})  we denote a  parton of type $i \in [-5,...,5]$ as $f_i$,
with $f_0 = g$ and $f_{\pm1,\pm2,\pm3,\pm4,\pm5} = 
\{d/\bar d,u/\bar u,s/\bar s,c/\bar c,b/\bar b\}$.  With a slight abuse 
of notation, we also use $f_i(x)$ to denote
the parton distribution function of parton $f_i$. 

The partonic
cross section ${\rm d} \hat\sigma_{f_i+e \to X}$ can be computed in QCD
perturbation theory as an expansion in the strong coupling constant
$\alpha_s$. We write 
\be {\rm d} \hat\sigma_{f_i+e \to X} = 
\dsh_{f_i}^\LO + \dsh_{f_i}^\NLO + \dsh_{f_i}^\NNLO
+ {\cal O}(\alpha_s^3).  
\ee
At leading order, electron-quark and electron-anti-quark scattering processes  
\begin{align}
\begin{split}
q(p_1) + e^-(p_2) &\to e^-(p_3) +q(p_4) \, , \\
\bar q(p_1) + e^-(p_2) &\to e^-(p_3) + \bar q(p_4) \, ,
\end{split}
\end{align}
contribute. 
For the purpose of computing QCD corrections, there is no difference
between these two processes and we focus on the electron-quark scattering.

To compute the partonic cross section of this process, 
we employ the notation that has been used in earlier papers on 
the \NSS~\cite{Caola:2017dug, Caola:2019nzf,Caola:2019pfz},  and define
\be
\begin{split} 
  \la \FLM(1_q,2_e,3_e,4_q)
      {\cal O}(\{1,2,3,4\}) \ra &\equiv ~{\cal N} \int \df{3} \df{4} 
\; (2 \pi)^{d} \delta^{(d)}(p_3 + p_4 - p_1 - p_2 )\;
  \\
& \quad \times
|\mathcal{A}^{\rm tree}|^2 (1_q,2_e,3_e,4_q)
\mathcal{O}(\{p_1,p_2,p_3,p_4\}) \, ,
\end{split} 
\label{eq2.5}
\ee
where ${\cal N}$ includes normalization and symmetry factors, $d=4-2\ep$ is
the space-time dimensionality,
\be
\df{i} = \frac{\d^{d-1}p_i}{(2\pi)^{d-1}2E_i} \; \theta(\Em-E_i) \, ,
\ee
is the phase-space volume of a  parton $f_i$, $\mathcal A^{\rm tree}$ is
the tree-level matrix element and ${\cal O}$ is a generic observable
that depends on momenta $p_{1,..,4}$.  $\Em$ is a sufficiently  large but
otherwise arbitrary\footnote{ More specifically, $\Em$ should be greater than or
  equal to the maximal energy that a final state parton can have
  according to the momentum conservation constraint.\label{ftn:emax}  } parameter  that provides 
an upper bound
on energies of individual partons; its role will become clear later.
We will also use the notation $\la \FLM(i,j,...) \ra_{\delta}$ to
indicate that the corresponding cross section is fully-differential
with respect to momenta that are shown as arguments of the function
$\FLM$. In this notation, the fully differential LO cross section for quark-electron scattering
reads
\be
2s\cdot \dsh_q^{\LO} = \la\FLM(1_q,2_e,3_e,4_q)\ra_\delta,
\ee
where $s=2 p_1 \cdot p_2 $ is the partonic center-of-mass energy squared.

We now discuss NLO QCD corrections. As  we already emphasized in 
Refs.~\cite{Caola:2017dug, Caola:2019nzf,Caola:2019pfz}, at this order in 
the $\alpha_s$ expansion,
our subtraction scheme is equivalent to 
the FKS one~\cite{Frixione:1995ms,Frixione:1997np}. In spite of that,  it is useful to discuss the NLO QCD
computation of DIS here, if only to develop a 
familiarity with our  notation. 
At NLO QCD, both the quark 
$q/\bar q+e \to q/\bar q+ e +g$ and the gluon $g+e \to q + \bar q +e$
channels contribute to the DIS cross section.
We consider the quark channel first, and start by discussing the real emission contribution.
For the sake 
of definiteness, we focus on the following process
\be
q(p_1) + e^-(p_2) \to e^-(p_3) +q(p_4)+g(p_5) \, .
\label{eq2.7}
\ee
In analogy with Eq.~\eqref{eq2.5}, we define
\begin{align}
\begin{split} 
&  \la \FLM(1_q,2_e,3_e,4_q,5_g)
      {\cal O}(\{1,2,3,4,5\}) \ra  \equiv {\cal N} \int \prod_{i=3}^{5} \df{i} \;
(2 \pi)^{d} \delta^{(d)}\left(p_1 + p_2 - \sum_{i=3}^5 p_i\right) \\
&\quad\times|\mathcal{A}^{\rm tree}|^2 (1_q,2_e,3_e,4_q,5_g)
\mathcal{O}(\{p_1,p_2,p_3,p_4,p_5\}).
\end{split} 
\end{align}
%
The scattering amplitude is singular when
the gluon $g_5$ is soft or when it is collinear to the incoming or
outgoing quark. Following our earlier work on the \NSS~\cite{Caola:2017dug, Caola:2019nzf,Caola:2019pfz}, 
we introduce operators $S_5$ and $C_{51},C_{54}$
to extract the leading soft and collinear behavior of scattering
amplitudes squared, and use these operators to isolate non-integrable singularities
in differential cross sections by systematically rewriting the identity
operator $\I$ as
\be
\I = S_5 + (I-S_5),\;\;\; \I = C_{51}+C_{54} + (\I - C_{51} - C_{54}). 
\ee
In both of the above equations, the first term describes a singular
contribution and the second is free from soft and collinear 
singularities. 

In the spirit of FKS subtraction \cite{Frixione:1995ms,Frixione:1997np}, we partition the phase space using
\be
1 = w^{51} + w^{54},
\label{eq:part}
\ee
where 
\be
w^{51} \equiv \frac{\rho_{54}}{\rho_{51}+\rho_{54}} \, , \quad w^{54} \equiv \frac{\rho_{51}}{\rho_{51}+\rho_{54}} \, , 
\ee
and 
\be
\rho_{5i} = 1 - \vec n_5 \cdot \vec n_i.
\label{eq2.10}
\ee
In Eq.~(\ref{eq2.10}) $\vec n_{i}$ are unit vectors that describe
directions of respective partons. The explicit form of the partition 
functions $w^{5i}$ in Eq.~\eqref{eq:part} is irrelevant, as long as they
have the following  property
\be
C_{5i} w^{5j} = \delta_{ij}, 
\ee
that leads to simplifications in the collinear limit. 
This allows us to write\footnote{To simplify the notation, from now on
we will not explicitly show 
electron arguments in the function $\FLM$.\label{ftn:drop}}
\be
\begin{split}
\label{eq2.9}
\la \FLM(1_q,4_q,5_g) \ra_\delta &=
\la  S_5 \, \FLM(1_q,4_q,5_g) \ra +  
\sum \limits_{i \in [1,4]} \la  C_{5i} (I-S_5)  \, \FLM(1_q,4_q,5_g) \ra
\\
& \quad + \sum \limits_{i \in [1,4]} \langle \hat{O}^{(i)}_{\rm NLO} w^{5i} \, 
\FLM(1_q,4_q,5_g) \rangle_\delta,
\end{split}
\ee
where 
\be
\hat{O}_{\rm NLO}^{(i)} \equiv (I - C_{5i})(I - S_5) \, .
\label{eq2.16}
\ee
In Eq.~(\ref{eq2.9}), the first term on the right hand side describes
the soft limit of the process Eq.~(\ref{eq2.7}), the second term
describes two soft-subtracted collinear limits and the last term describes
fully-regulated contributions that can be calculated in four
dimensions. 

We continue with the discussion of the different terms in
Eq.~(\ref{eq2.9}), starting with  the soft contribution.
We have
\be
\label{eq:nlo:q:soft_limit}
S_5 \, \FLM(1_q,4_q,5_g)  = 
2 g^2_{s,b} \Cf \frac{\rho_{14}}
{E_5^2 \rho_{51}^{\vphantom{2}}\rho_{54}^{\vphantom{2}}}
\FLM(1_q,4_q),
\ee
where $g_{s,b}$ is the bare QCD coupling and $\Cf=4/3$. Since
$\FLM(1_q,4_q)$ does not depend on $p_5$ anymore, we can 
integrate over the energy and angles of the unresolved gluon. 
We obtain 
\be
\int [{\rm d} f_5] \frac{1}{E_5^2} \frac{\rho_{14}}{\rho_{51}\rho_{54}} =
\frac{1}{\epsilon^2}
\left[ \frac{1}{8\pi^2} \frac{(4\pi)^\epsilon}{\Gamma(1 - \epsilon)} \right] 
\lp 4\Em^2\rp^{-\ep} \eta_{14}^{-\epsilon} K^{\vphantom \ep}_{14},
\label{eq:eik1}
\ee
where 
\be
K_{ij} = \left[\frac{\Gamma^2(1-\epsilon)}{\Gamma(1-2\epsilon)}\right] 
\eta_{ij}^{1+\epsilon} \, {}_2 F_1(1,1,1-\epsilon,1-\eta_{ij}), 
\ee
and $\eta_{ij} = \rho_{ij}/2 = (1-\cos\theta_{ij})/2$. 
This allows us to write
\be
\la S_5 \, \FLM(1_q,4_q,5_g)  \ra = \frac{2\Cf}{\epsilon^2} 
\lp\frac{4\Em^2}{\mu^2}\rp^{-\ep}
\qas
\bigg\langle  \eta_{14}^{-\epsilon} 
K_{14}^{\vphantom{-\epsilon}} \; \FLM(1_q,4_q) \bigg\rangle_\delta,
\label{eq2.18}
\ee
where we have introduced
\be
[\alpha_s] \equiv 
\frac{\alpha_s(\mu)}{2\pi} \frac{e^{\ep \gamma_E}}{\Gamma(1-\ep)}\,.
\ee
Note that at variance with cases of  color-singlet production and decay that were  studied
in Refs.~\cite{Caola:2019nzf,Caola:2019pfz}, the soft contribution depends non-trivially on
the angle between the two hard emittors.  

Next, we discuss the soft-subtracted collinear terms in Eq.~\eqref{eq2.9}.
We begin with the
term proportional to $C_{51}$ that describes the situation when 
the collinear gluon is emitted by an incoming quark. 
We parametrize the gluon energy as
\be
E_5 = (1-z) E_1,
\ee
and find 
\be
~~~~ \df{5} = \frac{\d\Omega_5^{d-1}}{2(2\pi)^{d-1}}
E_1^{2-2\ep} \d z (1-z)^{1-2\ep}
\theta\lp
z-\zm\rp,
\ee
where $\zm = 1-\Em/E_1$.  
The soft and collinear limits of the matrix element squared 
are well known. They read 
\bes
&
C_{51} |\mathcal A^{\rm tree}|^2(1_q,2_e,3_e,4_q,5_g) = 
\frac{g^2_{s,b}}{p_1\cdot p_5} \; \frac{P_{qq}(z)}{z}
|\mathcal A^{\rm tree}|^2(z\cdot1_q,2_e,3_e,4_q),
\\
&
C_{51}S_5 |\mathcal A^{\rm tree}|^2(1_q,2_e,3_e,4_q,5_g) = 
\frac{g^2_{s,b}}{p_1\cdot p_5} \; \frac{2\Cf}{1-z}
|\mathcal A^{\rm tree}|^2(1_q,2_e,3_e,4_q),
\end{split}
\label{eq2.24}
\ee
where $z\cdot 1_q$ denotes a quark with momentum $z\cdot p_1$ and
\be
P_{qq}(z) = \Cf\left[\frac{1+z^2}{1-z}-\ep(1-z)\right],
\label{eq:pqq}
\ee
is the splitting function for this limit. 

Since the matrix elements in Eq.~(\ref{eq2.24}) are independent
of the emission angles, we can integrate over them. The relevant integral reads 
\be
\int\frac{\d\Omega_5^{d-1}}{2(2\pi)^{d-1}}
\frac{1}{\rho_{5i}} = -\frac{4^{-\ep}}{\ep}
\left[\frac{1}{8\pi^2}\frac{(4\pi)^\ep}{\Gamma(1-\ep)}\right]
\frac{\Gamma^2(1-\ep)}{\Gamma(1-2\ep)}.
\label{eq2.26}
\ee
Putting everything together, we obtain 
\be
\begin{split} 
\la  C_{51} (I-S_5) \, \FLM(1_q,4_q,5_g) \ra &= 
-\frac{\qas}{\ep}\frac{\Gamma^2(1-\ep)}{\Gamma(1-2\ep)}
\lp\frac{4 E_1^2}{\mu^2}\rp^{-\ep}
\int\limits_{\zm}^{1} \d z (1-z)^{-2\ep}
\\
&\quad\times\la
P_{qq}(z)\frac{\FLM(z\cdot1_q,4_q)}{z} - 
\frac{2\Cf}{(1-z)}\FLM(1_q,4_q)
\ra_\delta.
\end{split}
\label{eq2.27}
\ee
We note that by construction $\Em \ge E_1$ (see footnote~\ref{ftn:emax}), 
so that $\zm \le 0$. This 
implies that 
$\FLM(z\cdot 1_q,4_q) = 0$ for $z\in[\zm,0]$ but the integration
of  the second term in angle brackets on the right hand side of Eq.~(\ref{eq2.27}) extends all the way to $\zm$. 
We isolate the term in  $P_{qq}$ that is singular in the $z \to 1$ limit,
\be
P_{qq}(z) = \frac{2\Cf}{1-z} + P_{qq,\rm reg}(z),~~~~
P_{qq,\rm reg}(z) = -\Cf\left[(1+z) +\ep(1-z)\right],
\ee
and write
\bes
&\la  C_{51} (I-S_5) \, \FLM(1_q,4_q,5_g) \ra  = -\frac{\qas}{\ep}\frac{\Gamma^2(1-\ep)}{\Gamma(1-2\ep)}
\lp\frac{4 E_1^2}{\mu^2}\rp^{-\ep}
\\
&\quad\times
\bigg\{
2\Cf\frac{(\Em/E_1)^{-2\ep}-1}{2\ep}\la\FLM(1_q,4_q)\ra_\delta  + \int\limits_{0}^{1} \d z 
\left[ 2\Cf \left[\frac{(1-z)^{-2\ep}}{1-z}\right]_+ \right.
\\
&\qquad
+\left.
(1-z)^{-2\ep} P_{qq,\rm reg}(z)\right]
\la\frac{\FLM(z\cdot1_q,4_q)}{z}\ra_\delta \bigg\} \, .
\end{split}
\label{eq:nlo_c51}
\ee
Note that in Refs.~\cite{Caola:2019nzf,Caola:2019pfz}, we have chosen  $\Em=E_1$
but we prefer  to keep $\Em$ generic in the current computation. Indeed, since final results are supposed to be
$\Em$-independent, the possibility to vary this parameter provides a useful 
check on the implementation of the subtraction formulas. The plus 
distribution in Eq.~\eqref{eq:nlo_c51} is defined as usual
\be
\int\limits_0^1 \d x \big[f(x)\big]_+ \cdot g(x) \equiv
\int\limits_0^1 \d x f(x) \big[g(x)-g(1)\big].
\ee

The discussion of the final-state collinear singularity, extracted by applying the operator $C_{54}$ to the 
matrix element squared,  is very similar. In this case we define 
\be
E_5 = (1-z)E_{45},~~~~ E_4 = z E_{45},
\label{eq2.30}
\ee
and repeat steps similar to the ones that led to Eq.~\eqref{eq:nlo_c51}. We obtain 
\begin{align}
\begin{split}
&\la  C_{54} (I-S_5) \, \FLM(1_q,4_q,5_g) \ra_\delta \\[5pt]
&= 
-\frac{\qas}{\ep}\frac{\Gamma^2(1-\ep)}{\Gamma(1-2\ep)} \bigg\langle
\bigg[
2\Cf\frac{(4\Em^2/\mu^2)^{-\ep}-(4E_4^2/\mu^2)^{-\ep}}{2\ep}
+\lp\frac{4E_4^2}{\mu^2}\rp^{-\ep}\\
&\quad \times \int\limits_0^1 \d z
\lp [z(1-z)]^{-2\ep} P_{qq}(z)-\frac{2\Cf}{(1-z)^{1+2\ep}}\rp
\bigg]\FLM(1_q,4_q)\bigg\rangle_\delta.
\label{eq:nlo_c54}
\end{split}
\end{align}
We note that further  details about final-state collinear splittings can be 
 found in the discussion of QCD corrections to  color-singlet decays,
see Ref.~\cite{Caola:2019pfz}.

To facilitate the $\ep$-expansion of Eq.~(\ref{eq:nlo_c54}), we write 
\bes
&2\Cf \left[\frac{(1-z)^{-2\ep}}{1-z}\right]_+
+
(1-z)^{-2\ep} P_{qq,\rm reg}(z)  = 
\PAP_{qq,R}^{(0)}(z) - \ep \mathcal P'_{qq}(z) + \mathcal O(\ep^2),
\\
&
-\int\limits_0^1 \d z
\lp [z(1-z)]^{-2\ep} P_{qq}(z)-\frac{2\Cf}{(1-z)^{1+2\ep}}\rp 
= \gamma_q + \ep \gamma'_q + \mathcal O(\ep^2).
\end{split}
\label{eq2.32}
\ee
The various splitting functions and anomalous dimensions are reported
in Appendix~\ref{sec:app_split}. We also define
\be
\Delta(E_i^2,E_j^2) = \frac{(4 E_i^2/\mu^2)^{-\ep}-(4 E_j^2/\mu^2)^{-\ep}}
{2\ep},
\label{eq2.33}
\ee
and write  the real contribution to the NLO cross section
Eq.~\eqref{eq2.9}
as follows 
\begin{align}
\begin{split}
&\la\FLM(1_q,4_q,5_g)\ra _\delta
= 
\frac{2\Cf}{\ep^2}\lp\frac{4\Em^2}{\mu^2}\rp^{-\ep}
\qas \big\langle \eta_{14}^{-\ep}K_{14} \FLM(1_q,4_q)\big\rangle_\delta
\\
&
\quad\quad+
\frac{\qas}{\ep}\frac{\Gamma^2(1-\ep)}{\Gamma(1-2\ep)} 
\Bigg\{
\bigg\langle \big[2\Cf\Delta(E_1^2,\Em^2)
+ 2\Cf\Delta(E_4^2,\Em^2)\big]
\FLM(1_q,4_q)
\bigg\rangle_\delta
\label{eq:nlor}
\\
&
\quad\qquad+
\lp\frac{4E_1^2}{\mu^2}\rp^{-\ep}
\int\limits_0^1\d z \left[- \PAP^{(0)}_{qq,R}(z)
+\ep \mathcal P'_{qq}(z)\right]
\la\frac{\FLM(z\cdot1_q,4_q)}{z}\ra_\delta
\\
&
\quad\qquad+
\lp\frac{4E_4^2}{\mu^2}\rp^{-\ep}
\left[\gamma_q + \ep\gamma'_q\right]\la\FLM(1_q,4_q)\ra_\delta
+\mathcal O(\ep^2)\Bigg\} \\
&
\quad\quad+
\sum \limits_{i \in [1,4]}^{} \langle \hat{O}^{(i)}_{\rm NLO} \, w^{5i} 
\; \FLM(1_q,4_q,5_g) \rangle_\delta.
\end{split}
\end{align}

As the next step, we consider virtual corrections. Using notation
analogous to Eq.~\eqref{eq2.5}, we define
\bes
&\la\FLV(1_q,2_e,3_e,4_q)\mathcal O(\{1,2,3,4\}) \ra \equiv
\mathcal N \int \df3\df4 \;
(2\pi)^d\delta^{(d)}(p_3+p_4-p_1-p_2)
\\
&\quad\times
2\Re\big[\mathcal A^{\rm tree}\mathcal A^{*,\rm1-loop}\big]
(1_q,2_e,3_e,4_q)\mathcal O(\{p_1,p_2,p_3,p_4\}). 
\end{split}
\ee
We employ  the Catani's representation~\cite{Catani:1998bh} for the 
renormalized amplitudes to write the NLO contribution
as\footnote{Also in this case, we do not show electron momenta 
in $\FLV$, see footnote~\ref{ftn:drop}.}
\bes
\la \FLV(1_q,4_q) \ra_\delta
= &
\lp\asontwopimu\rp
\la
I_{14}(\ep) 
\FLM(1_q,4_q) \ra_\delta + \langle \FLVf(1_q,4_q) \rangle_\delta\, ,
\label{eq:nlov}
\end{split}
\ee
where 
\be
I_{14}(\ep) =  -\frac{e^{\ep\gamma_E}}{\Gamma(1-\ep)}
e^{-\ep L_{14}} \left( \frac{2\Cf}{\ep^2}
 + \frac{2\gamma_q}{\ep} \right), 
\label{eq:i14}
 \ee
and
\be
L_{14} = \ln \frac{2(p_1 \cdot p_4)}{\mu^2}.
\ee
We note  that  $\gamma_q$ can be found  in Appendix~\ref{sec:app_split} 
(see Eq.~\eqref{eq:ga}) and 
that $\FLV^{\rm fin}$ is a finite remainder, free of any singularities.

To obtain the final result for the NLO QCD cross section, we combine
the real-emission contribution Eq.~\eqref{eq:nlor}
with virtual corrections Eq.~\eqref{eq:nlov} and the contribution that originates from the  
collinear renormalization of parton distribution functions
\be
2s\cdot\dsh^{\NLO}_{\rm PDF} = 
\asontwopimu
\int\limits_0^1\d z \PAP^{(0)}_{qq}(z) \la\frac{\FLM(z\cdot 1_q,4_q)}{z}
\ra_\delta,
\ee
where  $\PAP_{qq}^{(0)}$ is the Altarelli-Parisi splitting function, see 
Appendix~\ref{sec:app_split}. We find 
\bes
2s\cdot  \dsh_q^{\NLO} &=  \sum \limits_{i \in [1,4]}
  \langle \hat{O}^{(i)}_{\rm NLO} \; w^{5i} \; 
\FLM(1_q,4_q,5_g) \rangle_\delta + \langle \FLV^{\rm fin}(1_q,4_q) \rangle_\delta \\
&
+\asontwopimu\Bigg\{
\int\limits_0^1\d z
\bigg\langle
\left[\calP'_{qq}(z)+ \ln\lp\frac{4E_1^2}{\mu^2}\rp\PAP^{(0)}_{qq}(z)\right]
\frac{\FLM(z\cdot 1_q,4_q)}{z}\bigg\rangle_\delta
\\
&\quad+ 
\bigg\langle
\left[2\Cf\mathcal S^{\Em}_{14} + \gamma'_q\right]\FLM(1_q,4_q)
\bigg\rangle_\delta
 + \mathcal O(\ep)
\Bigg\},
\end{split}
\label{eq:nloq}
\ee
where the various (generalized) splitting functions and anomalous dimensions
are defined in Appendix~\ref{sec:app_split}. We have also defined
\bes
\mathcal S^{E}_{ij} &= 
\Li_2(1-\eta_{ij}) - \zeta_2 + \frac{\pi^2\lambda_{ij}}{2} + 
\frac{1}{2}\ln^2\lp\frac{E_i}{E_j}\rp 
-\ln(\eta_{ij})\ln\lp\frac{E_i E_j}{E^2}\rp
\\
&+
\frac{1}{2}\left[\frac{\gamma_i}{C_i}\ln\lp\frac{E_j\eta_{ij}}{E_i}\rp+
\frac{\gamma_j}{C_j}\ln\lp\frac{E_i\eta_{ij}}{E_j}\rp\right],
\end{split}
\label{eq:calS}
\ee
where $\gamma_i=\gamma_q(\gamma_g)$ and $C_i = \Cf(\Ca)$ if particle $i$
is a quark(gluon), and $\lambda_{ij} = 1$ if both particles $i$ and $j$ are 
in the initial or in the final state, and zero otherwise. We also
remind the reader that in our notation $\eta_{ij} = (1-\cos\theta_{ij})/2$,
where $\theta_{ij}$ is the angle between the directions of particle $i$ 
and $j$. 

Comparing Eq.~\eqref{eq:nloq} to similar  results for the production 
and decay  of a color singlet, 
considered in Refs.~\cite{Caola:2019nzf,Caola:2019pfz},
we note two main differences. First, Eq.~\eqref{eq:nloq} depends 
non-trivially on the relative angle between the incoming and
outgoing hard quarks. Second, subtraction terms in Eq.~\eqref{eq:nloq}
explicitly depend on $\Em$. This {\it explicit}  dependence is supposed to 
be canceled by 
an {\it implicit} dependence contained in the $\ONLO \FLM(1_q,4_q,5_g)$ terms. 
Checking the $\Em$-independence provides a useful cross-check
of the correctness of the implementation of Eq.~\eqref{eq:nloq} in a numerical program. 
Furthermore, we note that $\Em$  controls the amount of
(soft) subtractions; by varying $\Em$, we move  contributions from  the
regulated hard emission term $\ONLO \FLM(1_q,4_q,5_g)$ to  integrated
subtractions. In this sense, $\Em$  is closely related to the
so-called $\xi_{\rm cut}$ parameters of the FKS formalism, c.f. Refs.~\cite{Frixione:1995ms,Frixione:1997np}. 

In addition to the quark-electron scattering, at NLO we have to consider
the gluon-electron scattering
\be
g(p_1) + e^-(p_2) \to e^-(p_3) +q(p_4)+\qb(p_5).
\ee
The matrix element that describes this process is singular when the quark
or anti-quark becomes collinear to the incoming gluon. These singularities
are physically equivalent, so we find it convenient to treat both of them
at once. To this end, we introduce the following  
partitioning
\be
1 = w_g^{41} + w_g^{51},  \quad \textrm{with} \quad
w_g^{41} \equiv \frac{\rho_{51}}{\rho_{41}+\rho_{51}} \, , 
\quad w_g^{51} \equiv \frac{\rho_{41}}{\rho_{41}+\rho_{51}} \,,
\label{eq2.42}
\ee
and define
\bes
&\la\FLMg(1,4,5) \mathcal O(\{1,...,5\}) \ra
\equiv
\mathcal N \int \df3\df4\df5 \;
(2\pi)^d\delta^{(d)}(p_3+p_4+p_5-p_1-p_2)
\\
&\quad
\times
w^{41}\big[|\mathcal A|^2(1_g,2_e,3_e,4_q,5_\qb)|^2 + 
|\mathcal A|^2(1_g,2_e,3_e,5_q,4_\qb)|^2 \big]
\mathcal O(\{p_1,p_2,p_3,p_4,p_5\}). 
\end{split}
\label{eq:defFLMg}
\ee
This effectively remaps both the $gq$ and the $g\qb$ singularities onto
the $\vec p_1||\vec p_5$ collinear configuration. 
Since a final state quark does  not induce soft singularities, 
a  subtraction formula for the gluon channel is simpler than the formula 
for the quark channel. Repeating the same steps that led to
Eq.~\eqref{eq:nloq}, it is straightforward to obtain
\bes
2s\cdot\dsh_g^\NLO &= 
\langle 
(\I-C_{51}) \FLMg(1,4,5) \rangle_\delta \\
&+\asontwopimu
\int\limits_0^1\d z
\left[\calP'_{qg}(z)+ \ln\lp\frac{4E_1^2}{\mu^2}\rp\PAP^{(0)}_{qg}(z)\right]
\sum_{f\in[q,\qb]}\bigg\langle
\frac{\FLM(z\cdot 1_f,4_f)}{z}\bigg\rangle_\delta.
\end{split}
\ee

\section{NNLO calculation}
\label{sec:NNLO}
In this section, we discuss the calculation of NNLO QCD
corrections. Many details of the calculation are very similar to the
color-singlet production and decay cases discussed in
Refs.~\cite{Caola:2019nzf,Caola:2019pfz} and we do not repeat them
here. Rather, we skim through the derivation of the subtraction
formalism and concentrate on new features that arise in the DIS case.

As we already remarked in the previous section, the most important new
feature is the fact that hard partons are not back-to-back. As the
result, the integrated subtraction terms become functions of the
opening angle between these partons. The second new feature is that we
work with an arbitrary $\Em$.  As we explained in the previous
section, the $\Em$-independence of the final result arises through a
non-trivial interplay of subtractions and fully-regulated
contributions. As such, $\Em$ provides both a
powerful tool to check the correctness of the implementation of the
subtraction framework in a numerical program and also allows us to
shuffle contributions from numerical subtractions to
analytically-integrated subtraction terms.

We find it convenient to deal with quark- and gluon-initiated  processes separately. 
The primary reason for that is that  only the former ones contain  double-soft singularities, while in 
the latter case  the only genuine NNLO singularities are of the collinear type. We start
by discussing the  quark channel. 

\subsection{Quark channel}
We consider collision of an electron and a quark and  write the differential NNLO partonic cross section as
\be
\dsh^{\NNLO}_{q} = \dsh_q^{\rm VV} + 
\dsh_q^{\rm RV} + \dsh_q^{\rm RR} + \dsh_q^{\rm PDF},
\label{eq:nnloq_split}
\ee
where
\begin{itemize}
\item $\dsh_q^{\rm VV}$ is the double-virtual contribution, which requires
the one-loop squared and the two-loop amplitudes for the 
$q+e\to e + q$ process;
\item $\dsh_q^{\rm RV}$ is the real-virtual contribution, which requires
the one-loop amplitude for the $q+e\to e + q + g$ process;
\item $\dsh_q^{\rm RR}$ is the double-real contribution, which  requires
the tree
level amplitudes for the $q+e\to e + q + g + g$, 
$q + e \to e + q + Q + \bar Q$, with $Q\ne q$, 
and $q + e \to e + q + q + \bar q$ processes;
\item $\dsh_q^{\rm PDF}$ originates from  the collinear renormalization of parton distributions.
\end{itemize}
To efficiently manage the flavor structure and to follow  what is  commonly being done, we 
arrange the different partonic contributions into  \emph{non-singlet}
and \emph{singlet} pieces. We now briefly describe how this is done.

\begin{figure}
\centering
\includegraphics[width=0.8\textwidth]{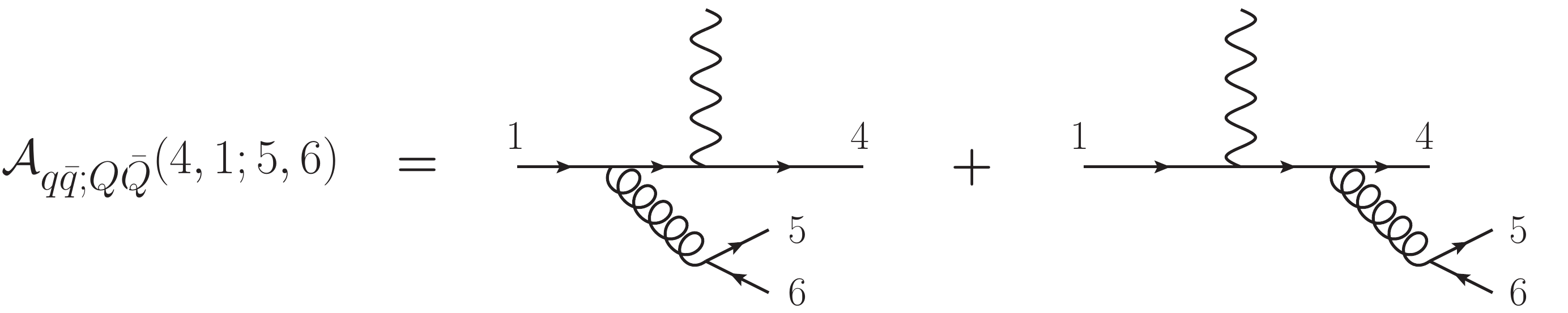}
\caption{Master amplitude for the $q(p_1)+e(p_2) \to 
e(p_3)+q(p_4)+Q(p_5)+\bar Q(p_6)$ processes. The electron line is
omitted. See text for details.}
\label{fig:amp}
\end{figure}

We consider the double-real contribution. Schematically, we write\footnote{We remind the reader that
momenta of electrons are not shown explicitly in formulas below.}
\bes
&
\dsh^{\rm RR}_q\sim\\
&
\int\prod_{i=3}^{6}\df{i}
\left\{
\sum_{q_i\ne q} |\mathcal A(1_q,4_q,5_{q_i},6_{\qb_i})|^2 + 
\frac{1}{2!}|\mathcal A(1_q,4_q,5_{q},6_{\qb})|^2 + 
\frac{1}{2!}|\mathcal A(1_q,4_q,5_{g},6_{g})|^2 \right\}.
\end{split}
\label{eq:rr_ch_q_ph}
\ee
We note that scattering amplitudes that involve an additional quark-anti-quark
pair can be constructed from the ``master''
amplitude $\mathcal A_{q\qb;Q\QB}$ shown in Fig.~\ref{fig:amp}. For example,  the  amplitude for the process $q(p_1)+e(p_2) \to
e(p_3)+q(p_4)+q'(p_5)+\qb'(p_6)$ with $q\ne q'$ is given
by\footnote{ Initial- and final-state crossings in $\mathcal
  A_{q\qb;Q\QB}$ are understood. Crossings are reflected in a 
$q$ vs. $\qb$ mismatch
in the flavor index between the subscripts and arguments of $\mathcal A$.}
\be
\mathcal A(1_q,4_q,5_{q'},6_{\qb'}) = \mathcal 
A_{q\qb;Q\QB}(4_q,1_q;5_{q'},6_{\qb'}) + 
\mathcal A_{q\qb;Q\QB}(5_{q'},6_{\qb'};4_q,1_q),
\ee
and the amplitude for the process $q(p_1)+e(p_2) \to 
e(p_3)+q(p_4)+q(p_5)+\qb(p_6)$ is then given by
\be
\mathcal A(1_q,4_q,5_q,6_{\qb}) = \mathcal A(1_q,4_q,5_{q'},6_{\qb'})+
\mathcal A(1_q,5_q,4_{q'},6_{\qb'}).
\ee
By systematically re-labeling partons, it is straightforward  to re-write 
$\dsh^{\rm RR}_{q}$ Eq.~\eqref{eq:rr_ch_q_ph} in terms of the amplitude 
$\mathcal A_{q\qb;Q\QB}$ as
\begin{align}
\begin{split}
\dsh^{\rm RR}_{q} &\sim 
\int \prod_{i=3}^{6}\df{i}
\Bigg\{
\sum_{q_i} \left[
|\mathcal A_{q\qb;Q\QB}(4_{q_i},6_{\qb_i};5_q,1_q)|^2
+|\mathcal A_{q\qb;Q\QB}(4_q,1_q;5_{q_i},6_{\qb_i})|^2\right]
\\
&\quad+\frac{1}{2!} |\mathcal A(1_q,4_q,5_g,6_g)|^2
+\sum_{q_i} 2\Re\big[
\mathcal A_{q\qb;Q\QB}(5_{q_i},6_{\qb_i};4_q,1_q)
\mathcal A^{*}_{q\qb;Q\QB}(4_q,1_q;5_{q_i},6_{\qb_i})\big]
\\
&\quad+\frac{2\Re}{2!}
\bigg[
\mathcal A_{q\qb;Q\QB}(5_q,1_q;4_q,6_\qb)
\left[
\mathcal A^*_{q\qb;Q\QB}(4_q,1_q;5_q,6_\qb)
+
\mathcal A^*_{q\qb;Q\QB}(4_q,6_\qb;5_q,1_q)
\right]
\\
&\qquad
+\mathcal A_{q\qb;Q\QB}(5_q,6_{\qb};4_q,1_\qb)
\left[
\mathcal A^*_{q\qb;Q\QB}(4_q,6_{\qb};5_q,1_q)+
\mathcal A^*_{q\qb;Q\QB}(5_q,1_q;4_q,6_{\qb})
\right]
\bigg]\Bigg\}.
\end{split}
\label{eq:ch_q_split}
\end{align}
We stress that, contrary to Eq.~\eqref{eq:rr_ch_q_ph}, 
the  sums in Eq.~(\ref{eq:ch_q_split}) run over 
\emph{all} quark flavors.

It is conventional to split the real-emission
processes  into non-singlet, singlet and the interference contributions. 
To that end, we write
\be
\dsh^{\rm RR}_{q} = 
\dsh^{\rm RR}_{q,\rm ns} +
\dsh^{\rm RR}_{q,\rm s} +
\dsh^{\rm RR}_{q,\rm int}.
\label{eq3.7}
\ee
The individual contributions in Eq.~(\ref{eq3.7}) are written as integrals of the
corresponding amplitudes squared 
\be
\dsh^{\rm RR}_{q,\rm ns,s,int} \sim \int\prod_{i=3}^{6}\df{i} \;
|\mathcal A_{\rm ns,s,int}(1,4,5,6)|^2.
\ee
These amplitudes read 
\bes
&|\mathcal A_{\rm ns}(1,4,5,6)|^2 =
\frac{1}{2!}|\mathcal A(1_q,4_q,5_g,6_g)|^2 + 
\sum_{q_i} |\mathcal A_{q\qb;Q\QB}(4_q,1_q;5_{q_i},6_{\qb_i})|^2
\\
&\quad
+\frac{2\Re}{2!}
\bigg[
\mathcal A_{q\qb;Q\QB}(5_q,1_q;4_q,6_\qb)
\left[
\mathcal A^*_{q\qb;Q\QB}(4_q,1_q;5_q,6_\qb)
+
\mathcal A^*_{q\qb;Q\QB}(4_q,6_\qb;5_q,1_q)
\right]
\\
&\qquad
+\mathcal A_{q\qb;Q\QB}(5_q,6_{\qb};4_q,1_\qb)
\left[
\mathcal A^*_{q\qb;Q\QB}(4_q,6_{\qb};5_q,1_q)+
\mathcal A^*_{q\qb;Q\QB}(5_q,1_q;4_q,6_{\qb})
\right] \bigg],
\\
&|\mathcal A_{\rm s}(1,4,5,6)|^2 =
\sum_{q_i}
|\mathcal A_{q\qb;Q\QB}(4_{q_i},6_{\qb_i};5_q,1_q)|^2,
\\
&
|\mathcal A_{\rm int}(1,4,5,6)|^2 =
\sum_{q_i} 2\Re\big[
\mathcal A_{q\qb;Q\QB}(5_{q_i},6_{\qb_i};4_q,1_q)
\mathcal A^{*}_{q\qb;Q\QB}(4_q,1_q;5_{q_i},6_{\qb_i})\big].
\end{split}
\label{eq3.8}
\ee

The different contributions have a distinct structure of infra-red and collinear singularities. 
The strongest singularities are present in the  \emph{non-singlet} contribution $|\mathcal A_{\rm ns}|^2$ which  exhibits 
non-vanishing  soft and collinear singular limits. More specifically, 
$|\mathcal A(1_q,4_q,5_g,6_g)|^2$ is singular if either one or both  gluons
are soft, or when they are   collinear to quarks $q_{1,4}$ or  to each other.
Other contributions to the non-singlet amplitude squared are less singular. 
For example,
$|\mathcal A_{q\qb;Q\QB}(4_q,1_q;5_{q_i},6_{\qb_i})|^2$  is singular when 
quarks $q_5$ and $q_6$ are both soft, or
when they are collinear to each other,
 or when they are simultaneously collinear to $q_1$ or $q_4$. 

The \emph{singlet} contribution $|\mathcal A_{\rm s}|^2$
only contains collinear singularities. 
Double-collinear singularities
 arise  when $q_5$ is collinear to $q_1$ or when $\qb_6$ 
is collinear to $q_4$. 
Triple-collinear singularities
arise when $q_1, q_5$ and $\qb_6$ are collinear or when $q_1, q_4$ and $q_5$ are collinear. 
The pure interference contribution
$|\mathcal A_{\rm int}|^2$ is {\it finite}. 
Furthermore, for photon-mediated DIS, that we consider in this paper,
the integral of this contribution vanishes due to Furry's theorem. 

Since double-virtual and real-virtual corrections only contribute to the non-singlet cross section, we write 
\be
\dsh^\NNLO_{q} = 
\dsh^\NNLO_{q,\rm ns} + 
\dsh^\NNLO_{q,\rm s} +
\dsh^\NNLO_{q,\rm int},  
\ee
where the three terms on the right hand side, defined as
\bes
&
\dsh^\NNLO_{q,\rm ns} = 
\dsh^{\rm RR}_{q,\rm ns} +
\dsh^{\rm RV}_{q} +
\dsh^{\rm VV}_{q} +
\dsh^{\rm PDF}_{q,\rm ns},\\
&
\dsh^\NNLO_{q,\rm s} = 
\dsh^{\rm RR}_{q,\rm s} +
\dsh^{\rm PDF}_{q,\rm s},\\
&
\dsh^\NNLO_{q,\rm int} = 
\dsh^{\rm RR}_{q,\rm int},
\end{split}
\label{eq:ds_q}
\ee
are separately finite. The collinear renormalization counterterms in 
Eq.~\eqref{eq:ds_q} are explicitly given by\footnote{In writing
 Eq.~\eqref{eq3.11}, we use the fact that 
in our case $\FLM(1_q,4_q)=
\FLM(1_\qb,4_\qb)$.}
\bes
\dsh_{q,\rm ns}^{\rm PDF} &= 
\left[\asontwopimu\right]^2
\dsh_q^\LO\otimes
\left[
\frac{\PAP^{(1)}_{qq,V}+\PAP^{(1)}_{q\qb,V}}{2\ep}
-\frac{
\PAP^{(0)}_{qq}\otimes\PAP^{(0)}_{qq}+\beta_0\PAP^{(0)}_{qq}}{2\ep^2}
\right]
\\
&+
\left[\asontwopimu\right]\dsh_q^\NLO\otimes \PAP^{(0)}_{qq};
\\
\dsh_{q,\rm s}^{\rm PDF} &= 
\left[\asontwopimu\right]^2
\dsh^\LO_q\otimes
\left[
\frac{\PAP^{(1)}_{qq,\rm s}}{2\ep} - \frac{
\PAP^{(0)}_{qg}\otimes\PAP^{(0)}_{gq}}{2\ep^2}
\right] + \left[\asontwopimu\right]\dsh^\NLO_{g}\otimes\PAP^{(0)}_{gq}.
\end{split}
\label{eq3.11}
\ee
As usual, the sign  ``$\otimes$'' stands for the convolution product and $\PAP^{(0/1)}$
are leading and next-to-leading order  Altarelli-Parisi splitting functions, see e.g.~\cite{Ellis:1991qj}. 
We report them in Appendix~\ref{sec:app_split} for convenience. 
The leading-order QCD $\beta$ function, which appears
in Eq.~(\ref{eq3.11}), reads 
\be
\beta_0 = \frac{11}{6}\Ca - \frac{2}{3}\tr\nf, 
\ee
where $\Ca = 3$, $\tr=1/2$ and $n_f$ is a number of massless quark flavors.

\subsubsection{NNLO corrections in the non-singlet channel: derivation}
\label{sec:ns_der}
The goal of this section is to  describe the calculation of  NNLO QCD corrections to neutral current DIS
in the non-singlet channel. Our goal in this discussion is two-fold. On the one hand, we aim to show  that 
many ingredients of the current computation 
are similar to cases of color-singlet production and decay discussed
in Refs.~\cite{Caola:2017dug, Caola:2019nzf,Caola:2019pfz} and, for this reason, can be directly
borrowed from these references. On the other hand, we want  to emphasize new elements
required for the construction of the nested subtraction scheme  when a process involves
color-charged initial- and final-state partons. 

We start by discussing the double-real contribution $\dsh^{\rm RR}_{q,\rm ns}$. 
It  contains double-soft singularities which  arise when partons $f_{5,6}$ 
become soft, 
$E_{5}\sim E_6 \to 0$.
Because of this, we find it convenient to order energies of partons $f_{5,6}$,  see Ref.~\cite{Caola:2017dug}.
Using the notation of Section~\ref{sec:LOandNLO}, 
we then define
\bes
&\la\FLMns(1,4,5,6)\mathcal O(\{1,...,6\})\ra \equiv
\mathcal N \int \prod_{i=3}^{6}\df{i}\;(2\pi)^{d}\delta^{(d)}\left(
p_1 + p_2 - \sum_{i=3}^{6}p_i\right)
\\
&\quad\times
\theta(E_5-E_6)\bigg[
|\mathcal A_{\rm ns}(1,4,5,6)|^2+
|\mathcal A_{\rm ns}(1,4,6,5)|^2\bigg]
\mathcal O(\{p_1,...,p_6\}),
\end{split}
\label{eq3.13}
\ee
so that
\be
2s\cdot \dsh^{\rm RR}_{\rm ns} = \la\FLMns(1,4,5,6)\ra_\delta.
\ee
To extract soft and collinear singularities from $\FLMns(1,4,5,6)$, 
we closely follow the procedure described in  Refs.~\cite{Caola:2017dug, Caola:2019nzf,Caola:2019pfz}.

First, we extract the double soft $E_5\sim E_6 \to 0$
singularity. Similar to  Refs.~\cite{Caola:2017dug, Caola:2019nzf,Caola:2019pfz},
we introduce an  operator $\SS$ that  extracts the leading soft  behavior of the matrix element, and sets
$E_{5,6}$ to zero in both the momentum-conserving $\delta$-function
and the observable $\mathcal O$ in Eq.~(\ref{eq3.13}). We write
\be
\la\FLMns(1,4,5,6)\ra_\delta = 
\la \SS \FLMns(1,4,5,6)\ra + \la (\I-\SS) \FLMns(1,4,5,6)\ra_\delta.
\label{eq3.14}
\ee
The second term on the r.h.s. of Eq.~\eqref{eq3.14} is free of double-soft
singularities. In the first term, partons $f_{5,6}$ completely decouple 
from the hard matrix element and any infra-red safe observable. Explicitly, we have 
\be
\la \SS \FLMns(1,4,5,6)\ra = 
\la \left[g_{s,b}^4\int\df5\df6 {\rm Eik}(1,4;5,6)\right] \FLM(1_q,4_q)\ra_\delta, 
\ee
where the function Eik(1,4;5,6) is the square of the double eikonal currents  computed e.g. in 
Ref.~\cite{Catani:1999ss}. Compared to the color-singlet production and decay cases
described in Refs.~\cite{Caola:2017dug, Caola:2019nzf,Caola:2019pfz}, the integral of ${\rm Eik}(1,4;5,6)$ 
depends on the relative angle between  directions of hard 
partons $f_{1,4}$. This integral was computed in Ref.~\cite{max_soft}
for a generic opening angle between $f_1$ and $f_4$, so we can directly
take the result from there.

The second term on the l.h.s. of Eq.~\eqref{eq3.14} is free from the double-soft
singularity, but still contains single-soft $E_6\to 0$ singularity.
To extract it, we use $\I = S_6 + (\I-S_6)$  and write 
\bes
  &\la (\I-\SS)\FLMns(1,4,5,6)\ra_\delta \\
  &=  \la S_6(\I-\SS)\FLMns(1,4,5,6)\ra  
+ \la (\I-S_6)(\I-\SS)\FLMns(1,4,5,6)\ra_\delta.
\end{split}
\label{eq3.18}
\ee

We deal with the first term on the right hand side  of Eq.~(\ref{eq3.18}) following the discussion of the  NLO computation, c.f. 
Eq.~\eqref{eq2.18}.
The only differences with respect to  that  case are $a)$ a more involved color structure
and  $b)$ the maximal allowed energy of $f_6$ is now $E_5$, because
of the energy ordering. Taking into account that the $S_6$ singularity is only
present if parton $f_6$ is a gluon, we obtain 
\bes
&\la S_6(\I-\SS)\FLMns(1,4,5,6)\ra \\[5pt]
&= 
\frac{\qas^2}{\ep^2}
\Bigg\langle\lp\frac{4 E_5^2}{\mu^2}\rp^{-\ep}
\bigg[
(2\Cf-\Ca)\eta_{41}^{-\ep} K_{14}+
\Ca\big[\eta_{51}^{-\ep} K_{15}+\eta_{54}^{-\ep} K_{54}\big]\bigg] \\
&\quad \times 
(\I-S_5)\FLM(1_q,4_q,5_g)
\bigg\rangle_\delta.
\end{split}
\label{eq3.19}
\ee
The right-hand side in Eq.~(\ref{eq3.19}) still contains unregulated 
singularities that arise when
$g_5$ is collinear to $q_1$ or $q_4$.  To extract them, we proceed as in the NLO 
case. To this end,  we again use the partition of unity  
shown in Eq.~\eqref{eq:part} and write
\be
\I = \sum_{i\in[1,4]}(\I-C_{5i})w^{5i} + \sum_i C_{5i},
\label{eq3.21}
\ee
where we used $C_{5i} w^{5j} = \delta_{ij}$. When Eq.~\eqref{eq3.21} is used in 
Eq.~\eqref{eq3.19}, the first term on the right hand side of  Eq.~\eqref{eq3.21} leads to a 
fully-regulated contribution, while the second term extracts the 
collinear divergences. Its integration over the unresolved phase space proceeds similar
to the NLO case except for two differences.
\begin{itemize}
\item When compared to NLO calculations, Eq.~\eqref{eq3.19} contains an 
additional $E_5^{-2\ep}$
factor. This leads to modified powers of $(1-z)^{-2\ep}$ in collinear limits
of differential cross sections. 
To efficiently
deal with this case, we find it convenient to define a generalised version
of Eq.~\eqref{eq2.32}. It reads
\bes
&\calP_{qq,Rn}(z) \equiv 2\Cf\left[\frac{(1-z)^{-n\ep}}{1-z}\right]_+
 +(1-z)^{-n\ep} P_{qq,\rm reg}(z),
\\
&\gamma_{qq}^{nk} \equiv
-\int\limits_0^1\d z \left[z^{-n\ep}(1-z)^{-k\ep}P_{qq}(z) 
-\frac{2\Cf}{(1-z)^{1+k\ep}}\right],
\end{split}
\label{eq3.22}
\ee
where $P_{qq}$ is the splitting function defined in Eq.~\eqref{eq:pqq}.

\item The pre-factor in Eq.~\eqref{eq3.19} leads to terms of the form
$C_{ij}\left[ \eta_{ij}^{-\ep} K_{ij}\right]$. This slightly changes the
angular integral that needs to be computed.
This point was already discussed in Refs.~\cite{Caola:2017dug, Caola:2019nzf,Caola:2019pfz}, 
and we refer
the reader to these references for details. 
\end{itemize}

Taking into account these two differences, and repeating steps explained
in the context of the NLO calculation, we obtain 
\bes
&\la C_{51}S_6(\I-\SS)\FLMns(1,4,5,6)\ra = 
-\frac{\qas^2}{\ep^3}
\int\limits_0^1\d z
\\
&\quad\times
\Bigg\langle
\Bigg[-\lp\frac{4 E_1^2}{\mu^2}\rp^{-2\ep}\calP_{qq,R4}(z)
+2\Cf \left[\frac{(4E_1^2/\mu^2)^{-2\ep}-(4\Em^2/\mu^2)^{-2\ep}}{4\ep}\right]
\delta(1-z)
\Bigg]
\\
&\qquad\times
\left[-2\Cf \frac{\Gamma^2(1-\ep)}{\Gamma(1-2\ep)}\eta_{41}^{-\ep} K_{14}
-\frac{\Ca}{2}\frac{\Gamma^4(1-\ep)\Gamma(1+\ep)}{\Gamma(1-3\ep)}
\right]\frac{\FLM(z\cdot 1_q,4_q)}{z}\Bigg\rangle_\delta,
\end{split}
\ee
for initial-state singularities, and 
\bes
&\la C_{54}S_6(\I-\SS)\FLMns(1,4,5,6)\ra \\
&= 
-\frac{\qas^2}{\ep^3}
\Bigg\langle
\Bigg[\lp\frac{4 E_3^2}{\mu^2}\rp^{-2\ep}\gamma_{qq}^{24}
+2\Cf \left[\frac{(4E_3^2/\mu^2)^{-2\ep}-(4\Em^2/\mu^2)^{-2\ep}}{4\ep}\right]
\Bigg]
\\
&\quad\times
\left[-2\Cf \frac{\Gamma^2(1-\ep)}{\Gamma(1-2\ep)}\eta_{41}^{-\ep} K_{14}
-\frac{\Ca}{2}\frac{\Gamma^4(1-\ep)\Gamma(1+\ep)}{\Gamma(1-3\ep)}
\right]\FLM(1_q,4_q)\Bigg\rangle_\delta,
\end{split}
\ee
for final-state ones. 

Combining the various terms discussed above, we arrive at the following formula
\be
\la \FLMns(1,4,5,6)\ra_\delta = 
\la(\I-S_6)(\I-\SS)\FLMns(1,4,5,6)\ra_\delta + ...
\label{eq3.25}
\ee
where ellipses stand for various contributions from which all soft and collinear singularities  have been extracted, as described above. 
However, the first term on the right  hand side  of Eq.~\eqref{eq3.25} still contains 
unregulated collinear singularities. To proceed with their 
extraction, we follow the FKS approach~\cite{Frixione:1995ms,Frixione:1997np} 
and its NNLO 
extension~\cite{czakonsub}, and partition the phase space
in the following way  
\be
1 = w^{51,61}+w^{54,64}+w^{51,64}+w^{54,61},
\label{eq3.26}
\ee
where $w^{5i,6j}$ are designed  to dampen all but a few collinear singularities.
More specifically, we ask that the damping factors
behave in the following way:
\be
C_{5i}w^{5j,6k}\propto \delta_{ij},~~~
C_{6i}w^{5j,6k}\propto \delta_{ik};~~~ i,j,k\in[1,4].
\ee
We also find it convenient to construct the functions $w^{5i,6j}$ 
in such a way that $w^{51,64}$ and $w^{54,61}$ vanish in the limit when  $f_5$ and $f_6$ become collinear to each
other  and that they are symmetric under the $5\leftrightarrow 6$ exchange. 
Apart from these requirements, the explicit form of the damping factors is
mostly immaterial. An explicit construction of these factors, valid also
for the DIS case,  
was discussed in Ref.~\cite{Caola:2017dug, Caola:2019nzf,Caola:2019pfz}.
We report it in Appendix~\ref{sec:app_damp} for convenience. 

In the ``double-collinear'' partitions $w^{51,64}$,
$w^{54,61}$, collinear singularities are effectively NLO-like.
In the ``triple-collinear'' partitions $w^{51,61}$ 
and $w^{54,64}$  genuine triple-collinear  singularities  occur when
partons $f_{5,6}$ become collinear to $f_{1}$ or $f_4$, respectively. 
However, since these triple-collinear limits can be reached in a variety of ways, it is useful
to introduce additional angular ordering
to isolate  physically-inequivalent configurations~\cite{czakonsub}. 
For the partition $w^{5i,6i}$, we write  
\bes
w^{5i,6i}  &=  w^{5i,6i} \left ( \theta\lp\rho_{6i}<\frac{\rho_{5i}}{2}\rp + 
 \theta\lp\frac{\rho_{5i}}{2}<\rho_{6i}<\rho_{5i}\rp + (5 \leftrightarrow 6) \right ) 
\\
&\equiv 
w^{5i,6i} \left ( \theta_i^{(a)}+\theta_i^{(b)}+\theta_i^{(c)}+\theta_i^{(d)} \right ).
\end{split}
\label{eq3.28}
\ee
We note in passing that the
angular ordering Eq.~(\ref{eq3.28})  can also be enforced by 
constructing appropriate damping factors
\cite{magnea}. Nevertheless, we find it practical to employ 
Eq.~(\ref{eq3.28}) to isolate and extract  collinear singularities
and to integrate the subtraction terms analytically.  
In particular, a phase space
parametrization  that naturally implements the sector decomposition as in 
Eq.~(\ref{eq3.28}) and that we employ 
in this paper 
can be found in Ref.~\cite{czakonsub}.

We  extract the remaining collinear 
singularities using Eqs.~(\ref{eq3.26},\ref{eq3.28}).
To this end, we follow  Refs.~\cite{Caola:2017dug, Caola:2019nzf,Caola:2019pfz} and write 
\bes
&\la(\I-S_6)(\I-\SS)\FLMns(1,4,5,6)\ra_\delta \\
&=
\la \FLMns^{s_r c_s}(1,4,5,6)\ra 
+ \la \FLMns^{s_r c_t}(1,4,5,6)\ra
+ \la \FLMns^{s_r c_r}(1,4,5,6)\ra_\delta.
\label{eq3.29}
\end{split}
\ee
The three terms on the right hand side in Eq.~(\ref{eq3.29}) are defined as follows: 
\begin{itemize} 
\item the soft-regulated single-collinear contribution $\FLMns^{s_r c_s}$ reads 
\begin{align}
\begin{split}
&\la\FLMns^{s_r c_s}(1,4,5,6)\ra \\
&= 
\sum_{(ij) \in [14,41]} 
\bigg\langle\bigg[C_{5i}\df5 + C_{6j}\df6\bigg]w^{5i,6j}
(\I-S_6)(\I-\SS)\times \FLMns(1,4,5,6)\bigg\rangle \\
&+ 
\sum_{i \in [1,4]} 
\bigg\langle
\bigg[\theta^{(a)}C_{6i}+\theta^{(b)}C_{56} + 
\theta^{(c)}C_{5i} + \theta^{(d)} C_{56}\bigg]
\label{eq3.30}
\\
&\qquad\times
\df5\df6 \; w^{5i,6i}
(\I-S_6)(\I-\SS)
\FLMns(1,4,5,6)\bigg\rangle,
\end{split}
\end{align}
\item the soft-regulated triple- and double-collinear contribution 
$\FLMns^{s_r c_s}$ reads 
\begin{align}
\begin{split}
&\la \FLMns^{s_r c_t}(1,4,5,6)\ra \\
&=  \sum_{i\in[1,4]} \bigg\langle
\bigg[ \theta^{(a)}(\I-C_{6i}) + \theta^{(b)}(\I-C_{65}) 
+ \theta^{(c)}(\I-C_{5i}) +\theta^{(d)}(\I-C_{65})\bigg]\\
&\quad\times\df5\df6 \; \CC_i w^{5i,6i} (\I-S_6)(\I-\SS)\FLMns(1,4,5,6)\bigg\rangle
\label{eq3.31}
\\
&-\sum_{(ij)\in[14,41]}
\la C_{5i}C_{6j} \df5\df6 w^{5i,6j}(\I-S_6)(\I-\SS)\FLMns(1,4,5,6)\ra,
\end{split}
\end{align}
\item  and, finally,  the fully-regulated term reads
\begin{align}
\begin{split}
&\la\FLMns^{s_r c_r}(1,4,5,6)\ra_\delta \\
&= 
\sum_{i\in[1,4]}
\bigg\langle\bigg[
\theta^{(a)}(\I-C_{6i}) + \theta^{(b)}(\I-C_{65}) + 
\theta^{(c)}(\I-C_{5i}) + \theta^{(d)}(\I-C_{65})\bigg] \\
&\quad\times \df5\df6 \; (\I-\CC_i)w^{5i,6i}(\I-S_6)(\I-\SS)\FLMns(1,4,5,6)
\bigg\rangle_\delta \label{eq3.32}\\
&
+ \sum_{(ij)\in[14,41]}
\bigg\langle\bigg[(\I-C_{5i})(\I-C_{6j})\bigg]\df5\df6 \;
(\I-S_6)(\I-\SS)\FLMns(1,4,5,6)\bigg\rangle_\delta.\hspace{-11pt}
\end{split}
\end{align}
\end{itemize} 
We note that in Eqs.~(\ref{eq3.30},\ref{eq3.31},\ref{eq3.32}) we used the
notation introduced in Refs.~\cite{Caola:2017dug, Caola:2019nzf,Caola:2019pfz}.
In particular, $\CC_i$ denote triple-collinear limits, and collinear operators
act on everything that appears to  the right of them. For example, 
by writing 
$C_{ij} \df j$  we indicate that the phase space for parton $j$ has to be taken
in the corresponding collinear limit, see Refs.~\cite{Caola:2017dug, Caola:2019nzf,Caola:2019pfz}  for details. 

A detailed discussion of double- and triple-collinear sectors,
both for initial- and final-state collinear singularities,
can be found in Refs.~\cite{Caola:2019nzf,Caola:2019pfz}. The fact that the discussion of these
limits does not change  is the consequence of the fact that collinear singularities  only 
depend on  color charges and types of external partons; as the result,  once the production and decay
of color singlets are understood, the description of similar limits in DIS naturally follows. 
For this reason  we do not repeat the discussion
of collinear limits per se but, instead,  illustrate {\it new} features that arise  in the DIS case by 
focusing on a few representative examples.

We start by discussing the 
contribution shown in Eq.~\eqref{eq3.31}. Compared to the cases studied
in Refs.~\cite{Caola:2019nzf,Caola:2019pfz}, the collinear limits now have an explicit $\Em$
dependence. For the triple-collinear limits, relevant results were obtained 
in  Ref.~\cite{maxtc}, and we refer the  reader to this reference
for details. For the double-collinear case, we need to 
evaluate\footnote{We note that to go from Eq.~\eqref{eq3.31}
to Eq.~\eqref{eq3.33}, we used $C_{5i}C_{6j}w^{5i,6j}=1$,
which follows from the definition of the $w^{5i,6j}$ damping factors.}
\bes
\mathcal{DC} = -\la \big[C_{51}C_{64}+C_{54}C_{61}\big]
\df5\df6  (\I-S_6)(\I-\SS)\FLMns(1,4,5,6)\ra.
\end{split}
\label{eq3.33}
\ee
In the non-singlet channel, $\mathcal{DC}$ is non-vanishing only if both
partons 5 and 6 are gluons. Schematically, Eq.~\eqref{eq3.33} reads
\bes
&\big[C_{51}C_{64}+C_{54}C_{61}\big]
\df5\df6 \; (\I-S_6)(\I-\SS)\FLMns(1,4,5,6) \\
&\sim \big[C_{51}C_{64}+C_{54}C_{61}\big]\df5\df6 \; \theta(E_5 - E_6) \theta(\Em - E_5)
(\I-S_6)(\I-\SS)
\\
&\quad\times
\left[
|\mathcal A^{\rm tree}|^2(1_q,2_e,3_e,4_q,5_g,6_g)
\df4 \; (2\pi)^d \delta^{(d)}\left( p_1+p_2-\sum_{i=3}^6 p_i\right)
\right].
\end{split}
\label{eq3.34}
\ee
We note that apart from the operator $S_6$  and the energy-ordering condition,
this expression is symmetric under the exchange $f_5\leftrightarrow f_6$.
Accounting
for that and renaming partons appropriately, it is easy to show that
Eq.~\eqref{eq3.34} can be written as
\bes
&\big[C_{51}C_{64}+C_{54}C_{61}\big]
\df5\df6 \; (\I-S_6)(\I-\SS)\FLMns(1,4,5,6) \\
&\sim \Big\{C^{s_r}_{51}C^{s_r}_{64}\theta_{E_{5}<\Em} \theta_{E_{6}<\Em}
+
\big[C_{51}S_5\theta_{E_6<E_5<\Em}\big]C^{s_r}_{64}+
\big[C_{64}S_6\theta_{E_5<E_6<\Em}\big]C^{s_r}_{51}\Big\}
\\
&\quad\times
\left[
|\mathcal A^{\rm tree}|^2(1_q,2_e,3_e,4_q,5_g,6_g)
\df4 \df5\df6 \; (2\pi)^d \delta^{(d)}\lp p_1+p_2-\sum_{i=3}^{6} p_i\rp
\right],
\end{split}
\label{eq3.35}
\ee
where we used the short-hand notations $C_{5(6)i}^{s_r} = C_{5(6)i}(\I-S_{5(6)})$
and $\theta_{X <  Y} = \theta(Y-X)$, $\theta_{X < Y < Z} = \theta(Y-X) \theta(Z-Y)$. 
We note that each of the three terms in the curly bracket in 
Eq.~\eqref{eq3.35} does not contain unregulated soft divergences. The
first term is just the product of two NLO-like contributions; for this reason, it can 
be immediately  read off from Eqs.~(\ref{eq:nlo_c51}, \ref{eq:nlo_c54}). 
In the second term, the $C_{51}S_5$ soft-collinear limit leads to
\bes
&\la C_{51} S_5\theta_{E_6<E_5<\Em}\df5 \; \FLMns(1,4,5,6)\ra \\[5pt]
&= -2\Cf\frac{\qas}{\ep} \lp\frac{4}{\mu^2}\rp^{-\ep} \la\int\limits_{E_6}^{\Em} \frac{\d E_5}{E_5^{1+2\ep}}\FLM(1_q,4_q,6_g)\ra \\
&= -2\Cf\frac{\qas}{\ep}\la \Delta(E_6^2,\Em^2)\FLM(1_q,4_q,6_g)\ra,
\end{split}
\label{eq3.36}
\ee
where the function  $\Delta$ is defined in Eq.~\eqref{eq2.33}. According to Eq.~\eqref{eq3.35},
we now have to take the soft-regulated collinear limit $C^{s_r}_{64}$ of
Eq.~\eqref{eq3.36}. For the term proportional to $\Em$, everything proceeds
as in the NLO case. The term proportional to $E_6$, however,  induces an extra
power of $(1-z)^{-2\ep}$ after performing the change of variables 
$E_{6} = (1-z)E_{64}$, see Eq.~\eqref{eq2.30}. As we already explained  when 
discussing single-soft singularities, this leads to a term proportional
to $\gamma_{qq}^{24}$, c.f.  Eq.~\eqref{eq3.22}. More precisely, we obtain
\bes
&\la C_{64}^{s_r}C_{51} S_5\theta_{E_6<E_5<\Em}\df5\df6 \FLMns(1,4,5,6)\ra
=-2\Cf\frac{\qas^2}{\ep^2}\\
&\quad\times
\bigg\langle
\bigg[\frac{(4E_4^2/\mu^2)^{-2\ep} \gamma_{qq}^{24}-
(4 E_4 \Em/\mu^2)^{-2\ep}\gamma_{qq}^{22}}{2\ep} 
+ 2\Cf \frac{\Delta^2(E_4^2,\Em^2)}{2}\bigg]\FLM(1_q,4_q)\bigg\rangle_\delta.\hspace{-10pt}
\end{split}
\ee
An analogous result can be found for the last term in the curly bracket 
of Eq.~\eqref{eq3.35} that describes the regulated initial-state radiation
and the soft final-state radiation.

The last double-real contribution that we need to discuss is 
the soft-regulated single-collinear term Eq.~\eqref{eq3.30}.
Once again, an in-depth discussion of this term can
be found in Refs.~\cite{Caola:2019nzf,Caola:2019pfz}, and we do not  repeat it here. Rather, we
illustrate the new features arising when considering the DIS process by focusing 
on the initial triple-collinear sectors $w^{51,61}\theta^{(a,c)}$. 
Once these cases are understood, generalization to other sectors does not pose conceptual challenges
and can be obtained  following the discussion in   
Refs.~\cite{Caola:2019nzf,Caola:2019pfz}. 
Schematically, we write
\bes
&\bigg[\theta^{(a)} C_{61} + \theta^{(c)} C_{51}\bigg]
\df5\df6 \; w^{51,61}(\I-S_6)(\I-\SS)\FLMns(1,4,5,6)\\
&\sim \bigg
[\theta\lp\rho_{61}-\frac{\rho_{51}}{2}\rp C_{61} + 
\theta\lp\rho_{51}-\frac{\rho_{61}}{2}\rp C_{51} 
\bigg]w^{51,61}\df5\df6 \; \theta_{E_5<E_6<\Em} 
\\
&\quad\times(\I-S_6)(\I-\SS)\left[
|\mathcal A^{\rm tree}|^2(1_q,2_e,3_e,4_q,5_g,6_g)
\df4 \; (2\pi)^d \delta^{(d)}\left( p_1+p_2-\sum_{i=3}^{6} p_i\right)
\right], \hspace{-15pt}
\end{split}
\label{eq3.38}
\ee
where we used the fact that in the non-singlet
channel the $C_{51}$ and $C_{61}$ limits are only singular
if both partons 5 and 6 are gluons. 
To proceed further, we follow the discussion of   the double-collinear
contribution. We use the symmetry of $\FLMns$ with respect to an interchange of $g_5$ and $g_6$
and re-write 
Eq.~\eqref{eq3.38} as
\bes
&\bigg[\theta^{(a)} C_{61} + \theta^{(c)} C_{51}\bigg]
\df5\df6 \; w^{51,61}(\I-S_6)(\I-\SS)\FLMns(1,4,5,6) \\
&\sim\Bigg\{
\bigg[C_{61}\big[\I \cdot \theta_{E_5<\Em} \theta_{E_6 < \Em} -S_6 \theta_{E_6<E_5<\Em} \big]
\bigg]
(\I-S_5)
\\
&\qquad
+ 
\big[C_{61}(\I-S_6)\big] \big[S_5\theta_{E_6<E_5<\Em}\big]\Bigg\}
w^{51,61}\df5\df6 \;
\theta\lp\rho_{61}<\frac{\rho_{51}}{2}\rp \\
&\quad\times
\left[
|\mathcal A^{\rm tree}|^2(1_q,2_e,3_e,4_q,5_g,6_g)
\df4 \; (2\pi)^d \delta^{(d)}\left( p_1+p_2-\sum_{i=3}^{6} p_i\right)
\right].
\label{eq3.39}
\end{split}
\ee
We now analyze the two terms in the curly bracket. We note that both of
them only contain collinear divergences. Indeed, the soft singularity is
always regulated, either by the  explicit subtraction, as in the first term, or by the  energy-ordering 
condition, as in the second one. 

It is easy to see that the structure of the first term in Eq.~(\ref{eq3.39})  is nearly identical to the 
NLO case except that in the soft-collinear limit   the upper bound on the energy
of the  parton $f_6$ is $E_5$ (contrary to $\Em$ in the NLO case). 
This observation allows us to immediately write the 
result for  this contribution following the discussion in  Section~\ref{sec:LOandNLO}. We find
\bes
&
\bigg[C_{61}\big[\I \cdot \theta_{E_5 < \Em} \theta_{E_6 < \Em} -S_6 \theta_{E_6<E_5<\Em} \big]
\bigg]
(\I-S_5)
\longrightarrow
\\
&
\quad \frac{\qas}{\ep}\int\limits_0^1\d z
\bigg\langle
\bigg[
-\lp\frac{4 E_1^2}{\mu^2}\rp^{-\ep}
\calP_{qq,R2}(z)+2\Cf \Delta(E_1^2,E_5^2)\delta(1-z)\bigg]\\
&
\qquad\times\wt^{51,61}_{6||1}
\lp\frac{\rho_{51}}{2}\rp^{-\ep}
(\I-S_5)
\frac{\FLM(z\cdot 1_q,4_q,5_g)}{z}\bigg\rangle.
\end{split}
\label{eq3.40}
\ee
 We note that  functions $\calP_{qq,R2}$ and $\Delta$,
that appear  in Eq.~\eqref{eq3.40},
are defined in  Eqs.~\eqref{eq3.22} and~\eqref{eq2.33}, respectively. 
Also,  with $\wt$,  we indicate the damping factor in the collinear limit,
i.e.
\be
\wt^{5i,6j}_{6||k} = \lim_{\rho_{6k}\to 0} w^{5i,6j}.
\ee
Finally, we note that the factor $(\rho_{51}/2)^{-\ep}$, which is not 
present in the NLO
case, arises from the ordering condition 
$\theta(\rho_{61}<\rho_{51}/2)$ in Eq.~\eqref{eq3.39}. 

Eq.~\eqref{eq3.40} still contains an unregulated collinear singularity that occurs
when $g_5$ and $q_1$ become collinear.\footnote{We note that the  collinear 
$\vec p_5 || \vec p_4$ singularity  is removed
by the $\wt^{51,61}_{6||1}$ damping factor.} We extract it in the usual way 
by inserting $\I = C_{51} + (\I-C_{51})$. The regularization of the term
proportional to $\calP_{qq}$ was explained in detail in Ref.~\cite{Caola:2019nzf}. 
The regularization of the term proportional to $\Delta$ is analogous to
what we just discussed for the double-collinear contribution. 

We then move to 
the second term in curly brackets of Eq.~\eqref{eq3.39},
which corresponds to the 
nested soft-collinear limit.   Since in the limit when $g_6$ is collinear to
$q_1$, the emission of the soft gluon $g_5$ can not resolve 
$g_6$ and $q_1$, we are allowed to write 
\bes
&S_5 C_{61}(\I-S_6) w^{51,61}|\mathcal A(1_q,2_e,3_e,4_q,5_g,6_g)|^2 \\
&= g^2_{s,b}\frac{2\Cf}{E_5^2}\frac{\rho_{14}}{\rho_{51}\rho_{54}}
\wt^{51,61}_{6||1}
C_{61}(\I-S_6) |\mathcal A(1_q,2_e,3_e,4_q,6_g)|^2.
\end{split}
\ee
Similar to the NLO case, 
the dependence on the momentum of gluon $g_5$ has disappeared from the hard matrix
element. However, a residual
dependence on $p_5$ in $\wt^{51,61}_{6||1}$ and in the pre-factor
$(\rho_{51}/2)^{-\ep}$ appeared. 
The  dependence on the kinematics of gluon $g_5$  is described by 
the following integral
\bes
&
g_{s,b}^2\int\limits_{E_6}^{\Em} \frac{\d E_5}{E_5^{1+2\ep}}
\frac{\d\Omega^{d-1}_{5}}{2(2\pi)^{d-1}}
\frac{\rho_{41}}{\rho_{51}\rho_{54}}
\lp\frac{\rho_{51}}{2}\rp^{-\ep}
\wt^{51,61}_{6||1} \\
&=
\frac{\qas}{\ep} \frac{(4\Em^2/\mu^2)^{-\ep}-(4E_6/\mu^2)^{-\ep}}{2\ep}
\la \lp\frac{\rho_{51}}{2}\rp^{-\ep}\wt^{51,61}_{6||1}\ra_{S_5},
\end{split}
\label{eq3.43}
\ee
where we defined
\be
\la \mathcal O \ra_{S_5} \equiv 
-\ep 
\left[\frac{1}{8\pi^2}\frac{(4\pi)^{\ep}}{\Gamma(1-\ep)}\right]^{-1}
\int\frac{\d\Omega_5^{d-1}}{2(2\pi)^{d-1}}
\frac{\rho_{41}}{\rho_{51}\rho_{54}}
\; 4^{\ep} \;
\mathcal O.
\label{eq3.44}
\ee
The contribution from the second term in the curly bracket in 
Eq.~\eqref{eq3.39} is then\footnote{We stress that in this equation
the energy of gluon 6 is only subject to the constrain $E_6<\Em$.}
\bes
&\big[C_{61}(\I-S_6)\big] \big[S_5\theta_{E_6<E_5<\Em}\big] \dots
\longrightarrow \\
& \quad \frac{\qas}{\ep} 
\Bigg\langle
 C_{61}(1-S_6)\df6 \; \frac{(4\Em^2/\mu^2)^{-\ep}-(4E_6/\mu^2)^{-\ep}}{2\ep}
 \FLM(1_q,4_q,6_g) \\
&\qquad\times
\left[
2\Cf
\la \lp\frac{\rho_{51}}{2}\rp^{-\ep}\wt^{51,61}_{6||1}\ra_{S_5}
\right]
\Bigg\rangle_\delta.
\end{split}
\ee
At this stage, we can treat the collinear singularity as before. We note
that the pre-factor  $E_6^{-2\ep}$  will lead to additional overall factors
of $(1-z)^{-\ep}$, as discussed above. Taking this into account and 
repeating  steps similar to what is done in the  NLO case, we obtain
\bes
&\big[C_{61}(\I-S_6)\big] \big[S_5\theta_{E_6<E_5<\Em}\big] \dots
\longrightarrow 
\\
&\quad
-\frac{\qas^2}{\ep^2}\int\limits_0^1\d z \;
\Bigg\langle
\bigg[
\frac{(4\Em E_1/\mu^2)^{-2\ep}\calP_{qq,R2}(z) 
- (4 E_1^2/\mu^2)^{-\ep}\calP_{qq,R4}(z)}{2\ep}
\\
&\qquad
+2\Cf \frac{\Delta^2(E_1^2,\Em^2)}{2}\delta(1-z)
\bigg]
\frac{\FLM(z\cdot 1,4)}{z}
\left[
2\Cf
\la \lp\frac{\rho_{51}}{2}\rp^{-\ep}\wt^{51,61}_{6||1}\ra_{S_5}
\right]
\Bigg\rangle_\delta.
\end{split}
\label{eq3.46}
\ee
This completes our discussion of the regularization of the triple-collinear
sectors $w^{51,61}\theta^{(a,c)}$.  Finally, we note that this procedure is generic
and that one can regulate all the remaining sectors following it. 

Before discussing the real-virtual and double-virtual contributions,
we comment on the explicit dependence of Eq.~\eqref{eq3.46} on the
damping factor $\wt$. In general, one expects that $1/\ep$ poles of the double-real
contribution are independent of the damping factors, since they 
do not  appear in any other part (double-virtual,
real-virtual etc.) of the calculation. 
Eq.~\eqref{eq3.46} seems to contradict this assertion. We will now show 
that this is not the case.
To this end,  we note 
that the sum over double-collinear partitions in Eq.~\eqref{eq3.30} leads to a 
contribution
which is almost identical to Eq.~\eqref{eq3.46}. The only differences 
in the double-collinear case with respect to Eq.~\eqref{eq3.46} are 
$(a)$ the damping factor $\wt^{54,61}_{6||1}$ instead of
$\wt^{51,61}_{6||1}$ and $(b)$ the lack of the $(\rho_{51}/2)^{-\ep}$ term, 
since we do not require angular ordering in the double-collinear partition. 
Taking the sum of the contributions that come from  double and triple collinear partitions, we obtain the following 
integral
\be
\la \Delta_{61} \ra_{S_5} \sim
-\ep\int\d\Omega^{d-1}_{5} \frac{\rho_{41}}{\rho_{51}\rho_{54}}
\Delta_{61},
~~~~~
\Delta_{61} =  \wt^{54,61}_{6||1}+\lp\frac{\rho_{51}}{2}\rp^{-\ep}
\wt^{51,61}_{6||1},
\label{eq3.47}
\ee
see Eq.~\eqref{eq3.44}. This amounts to replacing 
$(\rho_{51}/2)^{-\ep}\wt^{51,61}_{6||1}
\to \Delta_{61}$ in Eq.~\eqref{eq3.46}.
  The term $\la \Delta_{61} \ra_{S_5}$  enters the differential
cross section in the combination $\la \Delta_{61}\ra_{S_5}/\ep^2$, c.f. 
Eq.~\eqref{eq3.46}; this implies that we should {\it prove}  that the dependence of $\la \Delta_{61} \ra_{S_5}$
on the partitioning $w$  {\it only} starts  at  ${\cal O}(\ep^2)$. 

Since the integrand in Eq.~(\ref{eq3.47}) is singular in two collinear limits $\rho_{51} \to 0$ and $\rho_{54} \to 0$,
we need to regularize and extract these singularities.  
Following the (by now) standard practice, we write
\be
\la\Delta_{61}\ra_{S_5} = 
\la ( C_{51} + C_{54} ) \Delta_{61}\ra_{S_5} + 
\la (\I-C_{51} - C_{54})\Delta_{61}\ra_{S_5}.
\label{eq3.48}
\ee
In the first term on the r.h.s. of Eq.~\eqref{eq3.48} 
the dependence on the partitioning
is absent because
\be
C_{5i}  \wt^{5j,61}_{6||1} = \delta_{ij}.
\ee
The second term on the r.h.s. of Eq.~(\ref{eq3.48}) is fully regulated and we can expand  it in $\epsilon$.
We find
\be
\begin{split} 
&\la (\I - C_{51} - C_{54}) \Delta_{61}\ra_{S_5} \\
&=  
-\ep\int\d\Omega^{d-1}_{5} \left [  \frac{\rho_{41}}{\rho_{51}\rho_{54}} \left (
\wt^{54,61}_{6||1}+ \wt^{51,61}_{6||1}
\right ) - \frac{1}{\rho_{51}} - \frac{1}{\rho_{54}}
\right ] + {\cal O}(\ep^2).
\end{split} 
\label{eq3.50}
\ee
Since partitions are constructed to satisfy the completeness relation
\be
\wt^{54,61}_{6||1}+ \wt^{51,61}_{6||1} = 1,
\label{eq:completeness}
\ee
Eq.~(\ref{eq3.50}) immediately proves that $\la C_{51}\Delta_{61}\ra_{S_5}$ is independent of the partitioning through
${\cal O}(\ep)$; this translates into the
independence of $1/\ep$ poles on the partitioning in the physical
cross section.   On the contrary, there is a dependence 
on $\wt$ in the finite part
of Eq.~\eqref{eq3.46}, which is cancelled by an (implicit) partition dependence
in the fully-regulated contribution Eq.~\eqref{eq3.32}. 

Finally, we note that since  the damping
factors are  process-dependent, it is not possible to analytically
compute partitioning-dependent finite contributions  once and for
all. We do not view  this as a problem. Indeed, it is simple
to see that the $\Delta_{61}$ can be re-absorbed in a slightly different
definition of the collinear operator in Eq.~\eqref{eq3.32}. We do not
pursue this avenue further since,  in the DIS case,  we were able to compute
the required integrals $\la \Delta_{61}\ra_{S_5}$ analytically, 
using the damping factors given in Appendix~\ref{sec:app_damp}.
This computation is outlined in Appendix~\ref{sec:app_partint}.

We now briefly discuss the real-virtual and double-virtual contributions
to the partonic cross section Eq.~\eqref{eq:ds_q}. To discuss 
$\dsh_q^{\rm RV}$, we define
\bes
&\la \FLV(1_q,4_q,5_g)\mathcal O(\{1,...,5\})\ra \equiv
\mathcal N
\int\prod_{i=3}^{5}\df{i} \; (2\pi)^{d}\delta^{(d)}
\left(p_1 + p_2 - \sum_{i=3}^5 p_i\right) \\
&\quad\times 2\Re\left[\mathcal A^{\rm tree} \mathcal A^{*,\rm 1-loop}\right]
(1_q,2_e,3_e,4_q,5_g)\mathcal O(\{p_1,...,p_5\}),
\end{split}
\label{eq:rvq}
\ee
where $\mathcal A^{\rm 1-loop}$ is the UV-renormalized
one-loop amplitude. We then  write
\be
2s\cdot \dsh^{\rm RV}_{q} = \la \FLV(1_q,4_q,5_g)\ra_\delta. 
\ee
The function $\FLV(1_q,4_q,5_g)$ contains a soft singularity that arises when the energy
of $g_5$ vanishes,    and collinear  singularities that arise when  
the momenta of $g_5$ and $g_1$ or the momenta of
$g_5$ and $g_4$
are parallel.

We now sketch the procedure to extract
these singularities, and refer the reader to Refs.~\cite{Caola:2017dug, Caola:2019nzf,Caola:2019pfz} for additional details. 
To extract the soft singularity,  we write  $\I = S_5 + (\I-S_5)$. 
The soft limit of a generic massless one-loop scattering amplitude was studied
in Ref.~\cite{Catani:1999ss}. Adapting the general result to our case, we write 
\bes
&\la S_5 \FLV(1_q,4_q,5_g)\ra = 
2\Cf g_s^2 \mu^{2\ep}
\Bigg\langle
\frac{(p_1\cdot p_4)}{(p_1\cdot p_5)(p_4\cdot p_5)}\Bigg\{
\FLV(1_q,4_q)
\\
&\quad-\left[\Ca\frac{\qas}{\ep^2}
\frac{\Gamma^5(1-\ep)\Gamma^3(1+\ep)}
{\Gamma^2(1-2\ep)\Gamma(1+2\ep)}
\lp
\frac{\mu^2 (p_1\cdot p_4)}{2(p_1\cdot p_5)(p_4\cdot p_5)}
\rp^{\ep}+\asontwopimu\frac{\beta_0}{\ep} \right]\FLM(1_q,4_q)
\Bigg\}
\Bigg\rangle,\hspace{-10pt}
\end{split}
\label{eq3.51}
\ee
where the term proportional to $\beta_0$ appears because we deal with
UV-renormalized amplitudes. The structure of Eq.~\eqref{eq3.51} is similar to
the NLO case Eq.~\eqref{eq:nlo:q:soft_limit}. We only require a
generalization of the eikonal integral Eq.~\eqref{eq:eik1}. It reads
\bes
2^{-\ep}\int\df5 \frac{1}{E_5^{2+2\ep}}
\left[\frac{\rho_{14}}{\rho_{51}\rho_{54}}\right]^{1+\ep} &= 
\frac{(4\Em^2)^{-2\ep}}{\ep^2}
\left[\frac{1}{8\pi^2}\frac{(4\pi)^\ep}{\Gamma(1-\ep)}\right]
\eta_{14}^{-2\ep} \frac{\Gamma^2(1-2\ep)}{\Gamma(1-4\ep)}\\
&\quad\times
\left[
\frac{\eta_{14}^{1+3\ep}F_{21}(1+\ep,1+\ep,1-\ep,1-\eta_{14})}{4}
\right],
\end{split}
\ee
with 
\be
\eta^{1+3\ep}F_{21}(1+\ep,1+\ep,1-\ep,1-\eta) 
= 1 + 4\ep^2 \Li_2(1-\eta) + ... .
\ee
For a particular choice of $\Em$,
the regularization of collinear singularities was discussed in detail
in Refs.~\cite{Caola:2017dug, Caola:2019nzf,Caola:2019pfz}. Generalization to
arbitrary $\Em$ can easily be done following steps similar to the ones
discussed for the double-real contribution; for this reason
 we won't discuss it further.
At the end, the RV contribution is written as
\be
\la \FLV(1_q,4_q,5_g)\ra_\delta = 
\la \sum_{i\in[1,4]}\hat{\mathcal O}_{\rm NLO}^{(i)} w^{5i}
\FLV(1_q,4_q,5_g)\ra_\delta + {\rm terms~with~LO~kinematics},
\label{eq3.54}
\ee
with $\hat{\mathcal O}_{\rm ONLO}$ defined in Eq.~\eqref{eq2.16}.
In Eq.~\eqref{eq3.54}, all the implicit phase-space $1/\ep$ poles 
have been extracted.
To extract the explicit loop-integration poles in $\FLV$, 
we follow Ref.~\cite{Catani:1998bh}
and define
\be
\la \FLV(1_q,4_q,5_g)\ra_\delta = 
\qas \la I_{1_q 4_q 5_g}\FLM(1_q,4_q,5_g)\ra_\delta + 
\la \FLVf(1_q,4_q,5_g)\ra_\delta,
\label{eq3.58}
\ee
where 
\bes
I_{1_q 4_q 5_g} &=   e^{-\ep L_{14}}
(\Ca-2\Cf)\left[\frac{1}{\ep^2}+\frac{3}{2\ep}
\right]
\\
& -\left[e^{-\ep L_{15}} + 
\cos(\ep\pi)e^{-\ep L_{45}}
\right]
\left[
\Ca\lp\frac{1}{\ep^2}+\frac{3}{4\ep}\rp + \frac{\beta_0}{2\ep}
\right],
\end{split}
\label{eq3.56}
\ee
and $\FLVf$ is a finite $\mu$-dependent remainder. 
In Eq.~\eqref{eq3.56}, we use the 
notation
\be
L_{ij} = \ln\left (\frac{2 p_i\cdot p_j}{\mu^2}\right),
\label{eq:Lij}
\ee
with $p_i\cdot p_j > 0$. 

The last term we need to discuss is the double-virtual contribution. 
We define
\bes
&\la \FLVV(1_q,4_q)\mathcal O(\{1,...,4\}\ra \equiv
\mathcal N
\int \df3\df4 \; (2\pi)^{d}\delta^{(d)}\big(p_3+p_4-p_1-p_2\big)
\\
&\quad\times \bigg\{|\mathcal A^{\rm 1-loop}|^2(1_q,2_e,3_e,4_q) + 
2\Re\left[\mathcal A^{\rm tree} \mathcal A^{*,\rm 2-loop}\right]
(1_q,2_e,3_e,4_q)\bigg\} \\
&\quad \times
\mathcal O(\{p_1,p_2,p_3,p_4\}), 
\end{split}
\ee
so that
$2s\cdot \dsh_q^{\rm VV} = \la \FLVV(1_q,4_q)\ra_\delta$.
To extract  all $1/\ep$ poles, we use results of Ref.~\cite{Catani:1998bh}
and write
\bes
2s\cdot \dsh_q^{\rm VV} & = 
\lp\asontwopimu\rp^2\bigg\langle
\bigg[
\frac{I_{14}^2(\ep)}{2} - \frac{\beta_0}{\ep} I_{14}(\ep)+
\frac{e^{-\ep\gamma_E}\Gamma(1-2\ep)}{\Gamma(1-\ep)}\lp\frac{\beta_0}{\ep}+K\rp
I_{14}(2\ep)
\\
&\quad
+\frac{e^{\ep\gamma_E}}{\Gamma(1-\ep)}\frac{H_q}{\ep}\bigg]
\FLM(1_q,4_q)\bigg\rangle_\delta
+ \lp\asontwopimu\rp \bigg\langle I_{14}(\ep) 
\FLVf(1_q,4_q)\bigg\rangle_\delta 
\\
&
+
\left\langle \FLVVf(1_q,4_q)\right\rangle_\delta + 
\left\langle \FLVsqf(1_q,4_q)\right\rangle_\delta, \vphantom{\bigg]}
\end{split}
\label{eq3.61}
\ee
where $\FLVVf$ and $\FLVsqf$ are finite remainders, 
see Refs.~\cite{Caola:2017dug, Caola:2019nzf,Caola:2019pfz}
for more details.\footnote{We note that $\FLVVf$ explicitly depends on the
renormalization scale $\mu$.} In Eq.~\eqref{eq3.61}, we used the
$I_{14}$ defined in   Eq.~\eqref{eq:i14},
$K = \lp{67}/{18}-\zeta_2\rp\Ca - {10}/{9}\;\tr\nf$, and
\bes
H_q &= \Cf^2\lp\frac{\pi^2}{2}-6\zeta_3-\frac{3}{8}\rp + 
\Ca\Cf\lp\frac{245}{216}-\frac{23}{48}\pi^2+\frac{13}{2}\zeta_3\rp
+\Cf\nf\lp\frac{\pi^2}{24}-\frac{25}{108}\rp.
\end{split}
\ee
We combine  the double-real, real-virtual and double-virtual 
contributions described in this section with the PDF collinear renormalization
Eq.~\eqref{eq3.11} and obtain a  fully-regulated finite final result. 
We present it
in Section~\ref{sec:ns_res}. 

\subsubsection{NNLO corrections in the non-singlet channel: results}
\label{sec:ns_res}
The procedure outlined in the previous section allows us to rewrite the 
non-singlet NNLO differential cross section as
\be
\dsh^{\rm NNLO}_{q,\rm ns} = 
\dsh^{\rm NNLO}_{q,\rm ns,3j}
+\dsh^{\rm NNLO}_{q,\rm ns,2j}
+\dsh^{\rm NNLO}_{q,\rm ns,1j},
\label{eq:ns_fin_jet}
\ee
where the three terms on the r.h.s. are  individually finite and
integrable in four dimensions. 
The term $\dsh^{\NNLO}_{q,\rm ns,3j}$ requires tree-level 
amplitudes with up to two additional partons relative to the Born configuration. It only
receives contributions from double-real emission processes, and it is given by
\be
2s\cdot \dsh^{\NNLO}_{q,\rm ns,3j} = 
\la\FLMns^{s_r c_r}(1,4,5,6)\ra_\delta,
\ee
with $\la\FLMns^{s_r c_r}(1,4,5,6)\ra_\delta$ defined in Eq.~\eqref{eq3.32}.

The second term, $\dsh^{\NNLO}_{q,\rm ns, 2j}$, requires tree and loop
amplitudes with at most one additional parton relative to the Born configuration;
it can be written as
\begin{align}
& 2s\cdot \dsh^{\NNLO}_{q,\rm ns, 2j} = 
\la \sum_{i\in[1,4]} \hat{\mathcal O}_{\rm NLO}^{(i)} w^{5i}
\FLVf(1_q,4_q,5_g)\ra_\delta + \asontwopimu \Bigg\{
\int\limits_0^1\d z 
\sum_{i\in[1,4]} \Bigg\langle
\hat{\mathcal O}_{\rm NLO}^{(i)} w^{5i}
\nonumber\\
&\quad\times 
\bigg[\calP'_{qq}(z) 
+ \lp\ln\frac{4E_1^2}{\mu^2}-\tilde\Delta'_{61}\rp
\PAP^{(0)}_{qq}(z)\bigg]
\frac{\FLM(z\cdot1_q,4_q,5_g)}{z}\Bigg\rangle_\delta
\nonumber \\
& +
\Bigg\langle \sum_{i\in[1,4]}\hat{\mathcal O}_{\rm NLO}^{(i)} w^{5i}
\bigg[(2\Cf-\Ca)\mathcal S^{E_5}_{14} + 
\Ca\big(\mathcal S^{E_5}_{15}+
\mathcal S^{E_5}_{45}\big)+\gamma'_q+\gamma'_g 
\label{eq3.65}
 \\
&\quad +
\sum_{j\in[1,4,5]} \tilde\Delta'_{6j}
\lp \gamma_j + 2 C_j \ln\frac{E_5}{E_j}\rp
\bigg]\FLM(1_q,4_q,5_g)\Bigg\rangle_\delta 
\nonumber \\
&+
\gamma_{k_\perp,g}\sum_{i\in[1,4]}
\Bigg\langle
(\I-S_5)(\I-C_{51}-C_{54})
\left[r^{(i)}_{\mu}r^{(i)}_{\nu}-\frac{[-g_{\perp,\mu\nu}]}{2}\right] 
\wt^{5i,6i}_{6||5}
\FLM^{\mu\nu}(1_q,4_q,5_g)\Bigg\rangle_\delta
\Bigg\}, \nonumber
\end{align}
where the one-loop finite remainder $\FLVf(1_q,4_q,5_5)$ is defined in 
Eq.~\eqref{eq3.58}, $\hat{\mathcal O}_{\rm NLO}^{(i)}$ is defined
in Eq.~\eqref{eq2.16}, $w^{5i}$ are the damping factors discussed in
Section~\ref{sec:LOandNLO}, the function $\mathcal S^{E}_{ij}$ is 
defined in Eq.~\eqref{eq:calS} and all splitting functions and anomalous
dimensions are defined in Appendix~\ref{sec:app_split}. Similarly to the
NLO case, $\gamma_i = \gamma_q(\gamma_g)$ and $C_i =  \Cf(\Ca)$ if 
particle $i$ is a quark(gluon). The $\tilde\Delta'$ functions are
remnants of the damping factors and are defined as
\begin{gather}
\tilde\Delta'_{61} = -\wt^{51,61}_{6||1}\ln\lp\frac{\eta_{51}}{2}\rp,~~~
\tilde\Delta'_{64} = -\wt^{54,64}_{6||4}\ln\lp\frac{\eta_{54}}{2}\rp,
\label{eq:defDelta}
\\
\tilde\Delta'_{65} = -\sum_{i\in[1,4]}
\wt^{5i,6i}_{6||5}\ln\lp\frac{\eta_{5i}}{4(1-\eta_{5i})}\rp,
\nonumber
\end{gather}
with $\wt^{5i,6j}_{6||k} = \lim_{\eta_{6k}\to 0} w^{5i,6j}.$
Finally, $\FLM^{\mu\nu}$ is defined indirectly through the following 
equation
\be
\FLM(1_q,4_q,5_g) = \sum_{\lambda = \pm} \epsilon_{\mu}^{\lambda}(p_5)
\epsilon_{\mu}^{\lambda,*}(p_5)\FLM^{\mu\nu}(1_q,4_q,5_g).
\ee
The vectors $r_{\mu}^{(i)}$ in Eq.~\eqref{eq3.65}
are remnants of spin-correlations, and can be thought
of as a particular choice for the gluon polarization vector. Indeed, they
satisfy $r^{(i)}\cdot p_5 = 0$ and $r^{(i)}\cdot r^{(i)} = -1$. Their 
role in the subtraction framework and their explicit construction
is discussed in details in 
Refs.~\cite{Caola:2017dug, Caola:2019nzf,Caola:2019pfz}, and we refer the
reader to those references for more details. Here, we mention 
that if the momentum of gluon 5 is parametrized as
\be
p_5 = E_5(1,\sin\theta_{5i}\cos\varphi_5,\sin\theta_{5i}\sin\varphi_5,
\cos\theta_{5i}),
\ee
the
$r$ vector reads 
\be
r^{(i)} = (0,-\cos\theta_{5i}\cos\varphi_5,-\cos\theta_{5i}\sin\varphi_5,
\sin\theta_{5i}).
\label{eq:rdef}
\ee

The last term in Eq.~\eqref{eq:ns_fin_jet} that needs to be consider is
$\dsh^{\rm NNLO}_{q,\rm ns,1j}$. It describes the exclusive process
 $q+e \to q+e$, and 
only requires
tree-level and loop amplitudes
with Born-like kinematics. It reads\footnote{In writing this equation,
we use the fact that for neutral-current DIS one has
$\FLM(1_q,4_q) = \FLM(1_{\qb},4_{\qb})$.}
\begin{align}
&
2s\cdot\dsh^{\rm NNLO}_{q,\rm ns,1j} = 
\la\FLVVf(1_q,4_q)\ra_\delta + \la\FLVsqf(1_q,4_q)\ra_\delta 
\nonumber
\\
&
+ 
\asontwopimu\Bigg\{\int\limits_0^1\d z
\left[\calP'_{qq}(z) + \ln\lp\frac{4 E_1^2}{\mu^2}\rp\PAP^{(0)}_{qq}(z)\right]
\la\frac{\FLVf(z\cdot 1_q,4_q)}{z}\ra_\delta
\nonumber
\\
&
\quad+\bigg[2\Cf \mathcal S_{14}^{\Em} + \gamma'_q\bigg]
\big\langle \FLVf(1_q,4_q)\big\rangle_\delta\Bigg\}
+\lp\asontwopimu\rp^2\Bigg\{
\bigg\langle
\bigg[
\Delta_{\rm NS}(E_1,E_4,\Em,\eta_{14})
\label{eq3.69}
\\
&
\quad+\Cf\lp\delta_{k_\perp,g} \la r^\mu r^\nu\ra_{\rho_{5}}
- \delta_g \la \Delta_{65}\ra_{S_5}''
+\widetilde{\gamma_q}(E_4,\Em)
\la \Delta_{64}\ra_{S_5}''\rp\bigg]
\FLM(1_q,4_q)\bigg\rangle_\delta
\nonumber
\\
&
\quad+\int\limits_0^1\d z
\bigg\langle
\bigg[
\Cf\widetilde{\mathcal P}_{qq}(z,E_1,\Em)
\la \Delta_{61}\ra_{S_5}'' + 
\mathcal T_{\rm NS}(z,E_1,E_4,\Em,\eta_{14})\bigg]
\frac{\FLM(z\cdot 1_q,4_q)}{z}\bigg\rangle_{\delta}
\Bigg\}.
\nonumber
\end{align}
We note that finite parts of loop amplitudes have been defined in 
Eqs.~(\ref{eq:nlov},\ref{eq3.61}). The function $\mathcal S^E_{ij}$ 
can be found in Eq.~\eqref{eq:calS}. The terms $\la \Delta_{ij}\ra''$ 
and $\la r^\mu r^\nu\ra$ 
are the only contributions where the explicit dependence on the choice of 
partition functions appear. They are discussed in 
Appendix~\ref{sec:app_partint}, see Eqs.~(\ref{eq:D2def},\ref{eq:spint}).
They are multiplied by the generalized splitting function
\begin{align}
\begin{split}
\widetilde{\mathcal P}_{qq}(z,E_1,\Em) &= -\Cf \Bigg\{2\DD1(z)-(1+z)\ln(1-z) +\ln\lp\frac{E_1}{\Em}\rp \\
&\quad\times
\left[ 2\DD0(z)-(1+z) + \delta(1-z)\ln\lp\frac{E_1}{\Em}\rp \right]
\Bigg\},
\end{split}
\end{align}
and anomalous dimension
\be
\widetilde{\gamma}_q(E_4,\Em) = \Cf\left[ -\frac{7}{4} + \frac{3}{2}
\ln\lp\frac{E_4}{\Em}\rp - \ln^2\lp\frac{E_4}{\Em}\rp\right].
\ee
The two functions $\Delta_{\rm NS}$ and 
$\mathcal T_{\rm NS}$ that appear in Eq.~\eqref{eq3.69} contain the bulk of 
the NNLO 
(integrated) subtractions. They contain both standard and harmonic 
polylogarithms, and can be found in an ancillary file provided with
this submission. All the other splitting functions
and anomalous dimensions used in Eq.~\eqref{eq3.69} are reported in 
Appendix~\ref{sec:app_split}.

\subsubsection{NNLO corrections in the singlet channel}
We turn to the discussion of the singlet contribution to the cross section
$\dsh^{\NNLO}_{q,\rm s}$ defined in  Eq.~\eqref{eq:ds_q}.  The singlet channel is much simpler 
than the non-singlet one discussed previously. This is so because  $(a)$ it only receives
contributions from the double-real emission and the collinear renormalization of
PDFs and $(b)$ the singularity structure of the double-real contribution is
very simple. In particular, it does not contain soft singularities and no genuine final-state collinear
singularities.   Indeed, the matrix element squared $|\mathcal A_{\rm s}(1,4,5,6)|^2$ defined in Eq.~\eqref{eq3.8} is singular
when $\vec p_1||\vec p_5$ and/or when $\vec p_1||\vec p_5||\vec p_6$ or $\vec p_1||\vec p_5||\vec p_4$. Since
the two triple-collinear configurations are physically equivalent, we find
it convenient to treat both of them at once. To this end, we first introduce
a partition which is analogous to the one we used for computing NLO 
corrections in the gluon channel, c.f.  Eq.~\eqref{eq2.42}, and write 
\be
1 = w_{\rm s}^{41} + w_{\rm s}^{61},  \quad \textrm{with} \quad
w_{\rm s}^{41} \equiv \frac{\rho_{61}}{\rho_{41}+\rho_{61}} \, , 
\quad w_{\rm s}^{61} \equiv \frac{\rho_{41}}{\rho_{41}+\rho_{61}} \,.
\ee
Then, we use the symmetry of the amplitude and the phase space to write
\begin{align}
\begin{split}
&\mathcal N \int \prod_{i=3}^{6}\df{i} \; (2\pi)^{d}\delta^{(d)}\lp
\sum_{i=3}^{6}p_i - p_1 - p_2\rp
|\mathcal A_{\rm s}(1,4,5,6)|^2
\mathcal O(\{p_1,...,p_6\})\\
&= \mathcal N \int \prod_{i=3}^{6}\df{i} \; (2\pi)^{d}\delta^{(d)}\lp
\sum_{i=3}^{6}p_i - p_1 - p_2\rp
|{\mathcal A}^{156}_{\rm s}(1,4,5,6)|^2
\mathcal O(\{p_1,...,p_6\}),
\end{split}
\end{align}
with 
\be
|{\mathcal A}^{156}_{\rm s}(1,4,5,6)|^2 = 
w_{\rm s}^{41} \bigg[|{\mathcal A}_{\rm s}(1,4,5,6)|^2
+|{\mathcal A}_{\rm s}(1,6,5,4)|^2\bigg].
\ee
This manipulation effectively remaps both the $\vec p_1||\vec p_5||\vec p_4$ and the
$\vec p_1||\vec p_5||\vec p_6$ singularities of $|{\mathcal A}_{\rm s}(1,4,5,6)|^2$
into the single configuration $\vec p_1||\vec p_5||\vec p_6$ of 
$|{\mathcal A}^{156}_{\rm s}(1,4,5,6)|^2$, which is only singular if
$\vec  p_1||\vec p_5$ and/or $\vec p_1||\vec p_5||\vec p_6$. 

Since $|{\mathcal A}^{156}_{\rm s}(1,4,5,6)|^2$ does not contain any
soft singularity,  it is not necessary to order partons $f_5$ and $f_6$ in energy. 
 Nevertheless, we find it practical to treat
all contributions to the quark channel in the same way. We then define
\bes
&\la\FLMs(1,4,5,6)\mathcal O(\{1,...,6\})\ra \equiv
\mathcal N \int \prod_{i=3}^{6}\df{i} \; (2\pi)^{d}\delta^{(d)}\left(p_1 + p_2-
\sum_{i=3}^{6}p_i\right)
\\
&\quad
\times\theta(E_5-E_6)\bigg[
|\mathcal A^{156}_{\rm s}(1,4,5,6)|^2+
|\mathcal A^{156}_{\rm s}(1,4,6,5)|^2\bigg]
\mathcal O(\{p_1,...,p_6\}),
\end{split}
\ee
so that
\be
2s\cdot \dsh^{\rm RR}_{q,\rm s} = \la \FLMs(1,4,5,6)\ra_\delta. 
\ee

As we already mentioned, $\la \FLMs(1,4,5,6)\ra$ only contains initial state
double- ($\vec p_1||\vec p_5$) and triple- ($\vec p_1||\vec p_5||\vec p_6$) singularities.
Their extraction proceeds similarly to what we described in 
Section~\ref{sec:ns_der}, so we don't discuss it here and just 
 present the final results. Similar to the  non-singlet case, we
write
\be
\dsh^{\NNLO}_{q,\rm s} = 
\dsh^{\NNLO}_{q,\rm s,3j} + 
\dsh^{\NNLO}_{q,\rm s,2j} + 
\dsh^{\NNLO}_{q,\rm s,1j}.
\label{eq3.77}
\ee
The fully-regulated fully-resolved contribution now reads
\begin{align}
2s\cdot \dsh^{\NNLO}_{q,\rm s,3j}  &= 
\sum_{i\in[1,4]}
\bigg\langle\bigg[
\theta^{(a)}(\I-C_{6i}) + \theta^{(b)} I + 
\theta^{(c)}(\I-C_{5i}) + \theta^{(d)} I \bigg]\nonumber\\
&\quad\times
\df5\df6 \; (\I-\CC_i)w^{5i,6i}\FLMs(1,4,5,6)
\bigg\rangle_\delta \\
&+ 
\sum_{(ij)\in[14,41]}
\bigg\langle\bigg[(\I-C_{5i})(\I-C_{6j})\bigg]\df5\df6 \;
\FLMs(1,4,5,6)\bigg\rangle_\delta.
\nonumber
\end{align}
We note that the $\theta^{(b,d)}$ sectors don't require regularization
since there is no single $\vec p_5||\vec p_6$ collinear singularity. Because of this,
one could easily do away with the sectors. Nevertheless, as we have
already mentioned, we find it convenient to use the same parametrization
for both the singlet and non-singlet quark channel, so we keep the sectors for the singlet contributions. 

The second term on the r.h.s. of Eq.~\eqref{eq3.77} can be written as
\bes
2s\cdot \dsh^{\NNLO}_{q,\rm s,2j} &= \asontwopimu\int\limits_0^1\d z \; \Bigg\langle\lp I-C_{51}\rp
\left[\calP'_{gq}(z) + \lp\ln\frac{4E_1^2}{\mu^2} - 
\tilde \Delta'_{61}\rp \PAP^{(0)}_{gq}(z)\right] \\
&\quad\times
\frac{\FLMg(z\cdot1,4,5)}{z}\Bigg\rangle_\delta,
\end{split}
\ee
where $\FLMg$ and $\tilde\Delta'_{61}$  have been defined in 
Eq.~\eqref{eq:defFLMg} and Eq.~\eqref{eq:defDelta} respectively, while all 
the splitting
functions can be found in Appendix~\ref{sec:app_split}. Finally, the
fully-unresolved contribution reads
\be
2s\cdot\dsh^{\NNLO}_{q,\rm s,1j} = 
\lp\asontwopimu\rp^2\int\limits_0^1\d z \;\mathcal T_{\rm S}(E_1,z)
\sum_{f\in[q,\qb]}\la\frac{\FLM(z\cdot 1_f,4_f)}{z}\ra_\delta,
\ee
where the (universal) transition function $\mathcal T_{\rm S}$ is reported
in an accompanying ancillary file.

\subsection{Gluon channel}
In this section, we discuss NNLO QCD corrections in the gluon channel $e+ g \to e + X$. 
Similar to what we did in the quark case, c.f. Eq.~\eqref{eq:nnloq_split}, we write
\be
\dsh^{\NNLO}_g = \dsh^{\rm RV}_{g} + \dsh^{\rm RR}_{g} + \dsh^{\rm PDF}_g,
\label{eq4.1}
\ee
We note that, at variance with the quark channel, there are  no double-virtual corrections in this case. We now briefly
discuss the three terms on the right hand side  of Eq.~\eqref{eq4.1}. 

Employing   notation familiar from the discussion  of the quark channel,
we write the real-virtual contribution as
\be
2s\cdot\dsh^{\rm RV}_g = \la \FLV(1_g,4_q,5_\qb)\ra_\delta.
\ee
The function $\FLV(1_g,4_q,5_\qb)$  contains unregulated collinear 
singularities when a quark or an anti-quark becomes collinear to the incoming gluon. 
To consider both singularities at once, 
we proceed as we did in the  NLO case and define
\bes
&\la\FLVg(1,4,5)\mathcal O(\{1,...,5\})\ra 
\equiv \mathcal N \int\prod_{i=3}^{5}\df{i} \;
(2\pi)^d\delta^{(d)}\left(p_1 + p_2 -\sum_{i=3}^{5}p_i \right) \\
&\quad\times
w_g^{41}
2\Re
\bigg[\mathcal A^{\rm tree}
\mathcal A^{*,\rm 1-loop}(1_q,2_e,3_e,4_q,5_\qb) 
+ 
\mathcal A^{\rm tree}
\mathcal A^{*,\rm 1-loop}(1_q,2_e,3_e,5_q,4_\qb) \bigg] \\
&\quad\times \mathcal O(\{p_1,...,p_5\}),
\end{split}
\label{eq4.3}
\ee
where the damping factor $w^{41}_g$  is given in Eq.~\eqref{eq2.42} and
$\mathcal A^{\rm 1-loop}$ is UV-renormalized one-loop amplitude.
We note that $\FLVg$,  defined as in Eq.~\eqref{eq4.3},  is only singular
when $\vec p_5||\vec p_1$. Since there are no soft singularities, $\Em$ does
not play any role in the regularization procedure, which therefore follows the discussion in
Ref.~\cite{Caola:2019nzf}. We refer the reader
to that reference for details. The final result can be written as
\be
2s\cdot\dsh^{\rm RV}_g = 
\la (\I-C_{51})\FLVfg(1,4,5;\mu^2)\ra_\delta + {\rm terms~with~LO~kinematics}.
\label{eq4.4}
\ee
In Eq.~\eqref{eq4.4}, $\FLVfg$ is the finite one-loop remainder defined 
through
\be
\la\FLVg(1,4,5)\ra_\delta = \qas\la I_{1_g 4_q 5_q} \FLMg(1,4,5)\ra_\delta
+\la \FLVfg(1,4,5)\ra_\delta,
\ee
where 
\bes
& I_{1_g 4_q 5_q}  = 
\cos(\ep\pi)(\Ca-2\Cf)\left[\frac{1}{\ep^2}+\frac{3}{2\ep}
\right] e^{-\ep L_{45}}
\left[
\Ca\lp\frac{1}{\ep^2}+\frac{3}{4\ep}\rp + \frac{\beta_0}{2\ep}
\right] \sum \limits_{i=4,5}\; e^{-\ep L_{1i}}\;,
\end{split}
\ee
and $L_{ij}$ is defined in Eq.~\eqref{eq:Lij}.

We now discuss the double-real contribution $\dsh^{\rm RR}_g$.
Similar to the quark channel,
we write it as
\be
2s\cdot \dsh^{\rm RR}_{g} = \la\FLM(1_g,4_q,5_\qb,6_g)\ra_\delta. 
\ee
The matrix element for the process $g(p_1)+e(p_2) \to e(p_3) + q(p_4) + 
\qb(p_5) + g(p_6)$ is singular in the following kinematic configurations:
\begin{itemize}
\item $q$ or $\qb$ are collinear to the incoming gluon;
\item the outgoing gluon is collinear to the incoming one, or to the 
outgoing (anti)quark;
\item the outgoing gluon is soft.
\end{itemize}
 Similar to the case of  real-virtual corrections, the 
$\vec p_1||\vec p_4$ and $\vec  p_1||\vec p_5$ singularities are equivalent. We then define
\bes
&\la\FLMg(1,4,5,6)\mathcal O(\{1,...,6\})\ra \equiv 
\mathcal N \int\prod_{i=3}^6 \df{i} \; (2\pi)^d \delta^{(d)}
\left(p_1 + p_2 - \sum_{i=3}^{6} p_i\right) \\
&\quad\times
w^{41}_g
\bigg[
|\mathcal A|^2(1_g,2_e,3_e,4_q,5_\qb,6_g)
+
|\mathcal A|^2(1_g,2_e,3_e,5_\qb,4_q,6_g)\bigg]\mathcal O(\{p_1,...,p_6\}),
\end{split}
\ee
which is regular in the $\vec p_4||\vec p_1$ limit. The regularization of the
remaining singularities in the function $\FLMg$ proceeds similarly to what we discussed
in the case of the quark channels. There is only one main difference: since in this
case there are no double-soft singularities, we do not order  partons
$f_5$ and $f_6$ in energy. This slightly changes the construction of the subtraction terms, as described in 
Refs.~\cite{Caola:2019nzf,Caola:2019pfz}. Taking this into account and 
repeating steps similar to the ones sketched in Section~\ref{sec:ns_der} we
regulate all the singularities in $\FLMg(1,4,5,6)$. 

Finally, we consider the PDF collinear renormalization term. For the 
gluon channel, it reads
\bes
\dsh^{\rm PDF}_g &= 
\left[\asontwopimu\right]^2 \dsh^{\LO}_q\otimes 
\left[\frac{\PAP^{(1)}_{qg}}{2\ep} - 
\frac{\PAP^{(0)}_{qg}\otimes\PAP^{(0)}_{gg}
+\PAP^{(0)}_{qq}\otimes\PAP^{(0)}_{qg}}{2\ep^2}\right]+\\
&+\left[\asontwopimu\right]
\left[
\dsh^{\NLO}_{q}\otimes \PAP^{(0)}_{qg}
+\dsh^{\NLO}_{g}\otimes \PAP^{(0)}_{gg}\right].
\end{split}
\label{eq4.9}
\ee
All the relevant Altarelli-Parisi splitting functions are reported
in Appendix~\ref{sec:app_split}. We combine Eq.~\eqref{eq4.9} with the
regulated double-real and real-virtual contributions, and obtain 
\be
\dsh^\NNLO_g = \dsh^\NNLO_{g,\rm 3j} + 
\dsh^\NNLO_{g,\rm 2j} + 
\dsh^\NNLO_{g,\rm 1j}.
\label{eq4.11}
\ee
The fully-regulated fully-resolved contribution reads\footnote{We note
that sometimes the action of the $C_{ij}$ operators is zero. For example
$C_{54} \FLMg(1,4,5,6) = 0$, since the configuration where the two outgoing
quarks are collinear to each other is not singular in this channel. 
Nevertheless, we retain the symmetric notation of Eq.~\eqref{eq4.12} for 
convenience.}
\begin{align}
\begin{split}
2s\cdot \dsh^\NNLO_{g,\rm 3j} &= 
\sum_{i\in[1,4]}
\bigg\langle\bigg[
\theta^{(a)}(\I-C_{6i}) + \theta^{(b)}(\I-C_{65}) + 
\theta^{(c)}(\I-C_{5i})\\
&\quad
+ \theta^{(d)}(\I-C_{65})\bigg]
\df5\df6 (\I-\CC_i)w^{5i,6i}(\I-S_6)\FLMg(1,4,5,6)
\bigg\rangle_\delta
\label{eq4.12} \\
&+
 \sum_{(ij)\in[14,41]}
\bigg\langle\bigg[(\I-C_{5i})(\I-C_{6j})\bigg]\df5\df6
(\I-S_6)\FLMg(1,4,5,6)\bigg\rangle_\delta.
\end{split}
\end{align}
The second term on the r.h.s. of Eq.~\eqref{eq4.11} reads
\bes
&
2s\cdot \dsh^{\NNLO}_{g,\rm 2j} = 
\la \lp\I-C_{51}\rp
\FLVfg(1,4,5)\ra_\delta + 
\asontwopimu\Bigg\{
\int\limits_0^1\d z \sum_{i\in[1,4]}
\sum_{f\in[q,\qb]}
\Bigg\langle
\hat{\mathcal O}_{\rm NLO}^{(i)} w^{5i} \\
&\quad\qquad\times  
\bigg[\calP'_{qg}(z) 
+ \lp\ln\frac{4E_1^2}{\mu^2}-\tilde\Delta'_{61}\rp
\PAP^{(0)}_{qg}(z)\bigg]
\frac{\FLM(z\cdot1_f,4_f,5_g)}{z}\Bigg\rangle_\delta \hspace{-20pt}
\\
&\quad\quad+
\Bigg\langle
\lp \I-C_{51}\rp
\bigg[\calP'_{gg}(z) 
+ \lp\ln\frac{4E_1^2}{\mu^2}-\tilde\Delta'_{61}\rp
\PAP^{(0)}_{gg}(z)\bigg]
\frac{\FLMg(z\cdot1,4,5)}{z}\Bigg\rangle_\delta
\\
&\quad\quad+
\Bigg\langle \lp\I-C_{51}\rp
\bigg[(2\Cf-\Ca)\mathcal S^{\Em}_{45} + 
\Ca\big(\mathcal S^{\Em}_{14}+
\mathcal S^{\Em}_{15}\big)+2\gamma'_q
\\
&\quad\qquad
+\sum_{i\in[1,4,5]} \tilde\Delta'_{6i}
\lp \gamma_i + 2 C_i \ln\frac{\Em}{E_i}\rp
\bigg]\FLMg(1,4,5)\Bigg\rangle_\delta \Biggr\},
\end{split}
\label{eq4.13}
\ee
where $\FLMg$, $\tilde\Delta_{6i}$ and $\mathcal S^{\Em}_{ij}$ have
been defined in Eqs.~(\ref{eq:defFLMg},\ref{eq:defDelta},\ref{eq:calS}),
 respectively, while all the splitting functions and anomalous dimensions
can be found in Appendix~\ref{sec:app_split}. 
Finally, the fully-unresolved contribution to Eq.~\eqref{eq4.11} reads
\bes
2s\cdot \dsh^{\NNLO}_{g,\rm 1j} = 
\asontwopimu\int\limits_0^1\d z \left[
\calP'_{qg}(z) + \ln\lp\frac{4E_1^2}{\mu^2}\rp\PAP^{(0)}_{qg}(z)\right]
\sum_{f\in[q,\qb]}\la\frac{\FLVf(z\cdot 1_f,4_f)}{z}\ra_\delta
\\
+\lp\asontwopimu\rp^2\int\limits_0^1\d z 
\sum_{f\in[q,\qb]}\left\langle
\mathcal T_{g}(z,E_1,E_4,\Em,\eta_{14})\cdot
\frac{\FLM(z\cdot1_f,4_f)}{z}\ra_\delta,
\end{split}
\ee
where $\FLVf$ has been defined in Eq.~\eqref{eq:nlov}, the various
splitting functions can be found in Appendix~\ref{sec:app_split} and
$\mathcal T_{g}$ is reported in an ancillary file that accompanies this
paper.

\section{Validation of results}
\label{sec:results} 
In this section, we validate the results 
for the NNLO corrections to the deep-inelastic scattering process
$P + e \to e + X$ obtained with our subtraction scheme by comparing them to 
 analytic results for the NNLO DIS coefficient 
functions~\cite{Kazakov:1990fu,Zijlstra:1992qd,Moch:1999eb}, as implemented
in {\sc Hoppet}~\cite{Salam:2008qg,Cacciari:2015jma,Dreyer:2016oyx}.
Since the goal of this paper is to validate fully differential formulas for NNLO corrections to DIS,
rather than to perform   phenomenological
studies  of this process, we consider the simplest possible 
setup that allows for a thorough cross-check.
To this end, we only consider photon-induced neutral-current DIS. 
Furthermore, we only consider contributions proportional to either the
gluon or the up-quark PDF. In other words, we define the non-singlet
and singlet quark distributions as $q_{\rm s} = q_{\rm ns} = u$,
which is sufficient for validating the results presented in this paper.

We now describe  the setup of our computation. We consider photon-induced
DIS collisions with hadronic center-of-mass energy equal to $\sqrt{s} = 100~{\rm GeV}$,
and consider the total DIS cross section  where the momentum transfer from
an electron to a proton $q^2 = -Q^2$ is  restricted to the interval $10~{\rm GeV} < Q < 100~{\rm GeV}$.
We include contributions of 5 massless flavors (2 up, 3 down) in the final state. 
We always use the NNPDF3.0 NNLO set~\cite{Ball:2014uwa} 
as implemented in {\sc Lhapdf}~\cite{Buckley:2014ana} for both the parton
distribution functions and the strong coupling. 

In order to check
the scale dependence of our result, we set the renormalization and
factorization scales to $\mu_R = \mu_F = Q_{\rm max} = 100~{\rm GeV}$ instead
of a  more natural choice $\mu = Q$. In order to study the robustness
of our framework, we did not devise a specific parametrization for the
phase space of the underlying DIS process. Specifically, we did not 
use a phase space that naturally accommodates the    $t-$channel
vector boson exchange. Hence, our phase-space parametrization is clearly not optimal. 
We believe that by not optimizing it  we stress-test
the numerical performance of our subtraction scheme.
In general, we find that we can get per mill precision on the NNLO total
cross section, corresponding to a few percent precision on the NNLO coefficient, 
in a few hours  on an 8-core machine. 

We now presents our results. At LO, we obtain
\bes
\sigma^\LO_{\rm \NSS} = 1418.89(1)~\pb,~~~
\sigma^\LO_{\rm an} = 1418.89~\pb,
\end{split}
\ee
where the subscript indicates whether the result has
been obtained from our fully exclusive calculation (``\NSS'') 
or from the direct 
integration of the analytic coefficient functions (``an'') over $Q^2$ and $z$.
The Monte Carlo integration error for the former is shown in parentheses;
for the analytic case, this error is always negligible, so we don't show it here.

For the NLO corrections, we find
\bes
\sigma^\NLO_{{\rm \NSS},q} = 101.16(4)~\pb,~~~
\sigma^\NLO_{{\rm an},q} = 101.12~\pb,
\end{split}
\ee
and
\bes
\sigma^\NLO_{{\rm \NSS},g} = -297.90(1)~\pb, ~~~
\sigma^\NLO_{{\rm an},g} = -297.91~\pb.
\end{split}
\ee
We have explicitly checked that a similar level of agreement exists
 for different choices of the renormalization and factorization
scales $\mu$. We now move to the NNLO corrections. For the non-singlet
quark channel, we obtain
\be
\sigma^\NNLO_{{\rm \NSS,ns}} = \big[33.1(2) -2.18(1)\cdot\nf\big]~\pb,~~~
\sigma^\NNLO_{{\rm an,ns}} = \big[33.1 -2.17\cdot\nf\big]~\pb.
\ee
For the singlet channel, we obtain
\be
\sigma^\NNLO_{{\rm \NSS,s}} = 9.19(2)~\pb,~~~
\sigma^\NNLO_{{\rm an,s}} = 9.18~\pb,
\ee
where $n_u$ and $n_d$ are the number of up and down quarks, respectively. 
Finally, for the gluon channel we find
\be
\sigma^\NNLO_{{\rm \NSS,g}} = -142.4(4)~\pb,~~~
\sigma^\NNLO_{{\rm an,g}} = -142.7~\pb.
\ee
It follows from the above results that we can
compute  the  NNLO DIS coefficients with a few per mill precision, and that the agreement 
between numerical results and analytical predictions is excellent.
We have checked that this also holds true for other values of
the factorization and renormalization scales. As we explained in 
Sections~\ref{sec:LOandNLO} and~\ref{sec:NNLO}, our framework contains a parameter
$\Em$ which allows us to control the amount of (soft) subtraction. As such,
one can view this as a prototype for a  $\xi_{\rm cut}$ parameter 
in the FKS formalism~\cite{Frixione:1995ms,Frixione:1997np}. 
We have explicitly checked that our results are $\Em$-independent.

Finally, we note that we  performed other checks by splitting  numerical
and analytic results into contributions of individual color factors. 
This allows us to cross-check subtle interference effects, which are 
color-suppressed and, hence, largely invisible in the {\it full} result for NNLO coefficients. 
We have found good agreement between numerical and analytic results for all such cases
as well. 

\section{Conclusion}
\label{sec:conclusion}
In this paper, we presented analytic results for NNLO QCD 
corrections to deep-inelastic scattering within the nested soft-collinear
subtraction scheme introduced by some of us in Ref.~\cite{Caola:2017dug}. 
These results allow us to
extend the nested subtraction scheme to processes involving partons both in the
initial and in the final state. We have validated our calculation 
by computing NNLO QCD corrections to inclusive neutral-current DIS  and comparing 
them against predictions obtained from a direct integration of 
analytic DIS coefficient functions. We found that despite a sub-optimal
parametrization of the DIS phase space in the numerical routines, our formalism
performed well and allowed us to check individual NNLO coefficients to a few per mill precision. 

Apart from their relevance for processes like DIS or vector
boson fusion in the factorized approximation, the results presented here
constitute the last building block for applying the nested subtraction scheme to generic collider
processes. Indeed, the nested subtraction scheme has been previously formulated for processes
involving two hard partons both in the initial~\cite{Caola:2019nzf} and
in the final~\cite{Caola:2019pfz} state. Since at NNLO the structure of
infrared singularities is basically dipole-like, 
those results combined  with the ones presented 
in this paper  provide all the necessary building blocks to deal with arbitrary collider
processes.

In practice, there  are still two small issues that must
be confronted when dealing with higher multiplicity reactions. First, the
framework, as currently formulated, involves some partitioning-dependent contributions
that must be dealt with in an efficient way, see the discussion 
around Eq.~\eqref{eq3.50}.  We are confident that this issue
can be dealt with by using  a small modification of the   subtraction operators. 
Second, for processes involving 4 or more partons, non-trivial color correlations appear.
Although we have not studied such effects in detail yet,  we do not anticipate that they 
would prevent us from extending the nested subtraction scheme to generic processes.
We leave the investigation of these issues to the future. 

\section*{Acknowledgements}
We would like to thank Arnd Behring for providing independent values
of the analytic results for the DIS coefficient functions. We are
grateful to Maximilian Delto for the discussion of triple-collinear
limits.  F.C. would like to thank TTP KIT for hospitality during the
final stages of this work.  The research of K.A. was supported by
Karlsruhe School of Particle and Astroparticle Physics (KSETA).  The
research of K.A. and K.M. was partially supported by the Deutsche
Forschungsgemeinschaft (DFG, German Research Foundation) under grant
396021762 - TRR 257. The research of F.C. was partially supported by
the ERC Starting Grant 804394 {\sc hipQCD}.  

\appendix 

\section{Splitting functions and anomalous dimensions}
\label{sec:app_split}
In this section, we collect results for the various (generalized) splitting
functions and anomalous dimensions used in our calculation. 
We start by listing the Altarelli-Parisi splitting functions. At LO, they
read (see e.g.~\cite{Ellis:1991qj})
\bes
&\PAP^{(0)}_{qq}(z) = \Cf \big[2\DD0(z)-(1+z)\big] + \gamma_q \delta(1-z),\vphantom{\bigg]}
\\
&\PAP^{(0)}_{qg}(z) = \tr \big[z^2+(1-z^2)\big],\vphantom{\bigg]}
\\
&\PAP^{(0)}_{gq}(z) = \Cf\left[\frac{1+(1-z)^2}{z}\right],
\\
&\PAP^{(0)}_{gg}(z) = 2\Ca\left[\DD0(z) + \frac{1}{z} + z(1-z)-2\right]
+\gamma_g\delta(1-z),
\label{eq:PAP0}
\end{split}
\ee
where we defined 
\be
\DD{i}(z) = \left[\frac{\ln^i(1-z)}{1-z}\right]_+.
\ee
We also define
\be
\PAP_{qq,R}^{(0)}(z)  = \PAP_{qq}^{(0)}(z) - \gamma_q \delta(1-z); \;\;\;\;\;\;\PAP_{gg,R}^{(0)}(z)  = \PAP_{gg}^{(0)}(z) - \gamma_g \delta(1-z).
\ee
The LO anomalous dimensions in Eq.~\eqref{eq:PAP0} are defined as
\be
\gamma_q = \frac{3}{2}\Cf,~~~~
\gamma_g = \beta_0 = \frac{11}{6}\Ca - \frac{2}{3}\tr\nf.
\label{eq:ga}
\ee

For the NNLO calculation, we also require the following NLO
Altarelli-Parisi splitting functions (see e.g.~\cite{Ellis:1991qj})
\begin{align}
\PAP^{(1)}_{qq,V}(z) &= 
\Ca \Cf \Bigg[\left(\frac{67}{9}-\frac{\pi^2}{3}\right)
   \DD0(z)+\left(-3 \zeta_3+\frac{17}{24}+\frac{11 \pi ^2}{18}\right) 
\delta (1-z)
\nonumber
\\
&\quad+
\frac{\left(1+z^2\right) \ln ^2(z)}{2 (1-z)}+
\frac{\left(5 z^2+17\right) \ln
   (z)}{6 (1-z)}+\frac{53-187z}{18}
+(1+z) \zeta_2
\bigg]
\nonumber
\\
&-
\Cf  \tr \nf \bigg[\frac{20}{9}
   \DD0(z)+\left(\frac{1}{6}+\frac{2 \pi ^2}{9}\right) \delta (1-z)+\frac{2
   \left(1+z^2\right) \ln (z)}{3 (1-z)}+\frac{4 (1-z)}{3}
\nonumber
\\
&\quad
-\frac{10
   (z+1)}{9}\bigg]
+\Cf^2 \bigg[\left(6 \zeta_3+\frac{3}{8}-\frac{\pi
   ^2}{2}\right) \delta (1-z)+
-\frac{1}{2}(1+z) \ln^2(z)
\label{eq:PAP1}
\\
&\quad
+\frac{\left(2 z^2-2 z-3\right) \ln (z)}{1-z}-\frac{2
   \left(1+z^2\right) \ln (1-z) \ln (z)}{1-z}-5(1-z)\bigg],
\nonumber
\\
\PAP^{(1)}_{q\qb,V}(z) &= 
\Cf \left(\Cf-\frac{\Ca}{2}\right) \bigg[
\frac{1+z^2}{1-z}\left(\ln ^2(z)-4 \Li_2(-z)-4 \ln (z+1) \ln (z)
-\frac{\pi^2}{3}\right)
\nonumber
\\
&\quad
+4 (1-z)+2 (z+1) \ln (z)\bigg],
\nonumber
\\
\PAP^{(1)}_{qq,\rm s}(z) &= 
\Cf \tr \bigg[-\frac{56 z^2}{9}+\left(\frac{8 z^2}{3}+5 z+1\right) \ln
   (z)+6 z+\frac{20}{9 z}-(z+1) \ln ^2(z)-2\bigg];
\nonumber
\end{align}
and the  convolutions
\begin{align}
\left[\PAP^{(0)}_{qq}\otimes\PAP^{(0)}_{qq}\right](z) &=
\Cf^2 \bigg[6 \DD0(z)+8 \DD1(z)+\left(\frac{9}{4}-\frac{2 \pi
   ^2}{3}\right) \delta (1-z)
\nonumber
\\
&\quad
+\frac{\left(3 z^2+1\right) \ln (z)}{z-1}+2 (z-1)-3 z-4
   (z+1) \ln (1-z)-3\bigg],
\nonumber
\\
\left[\PAP^{(0)}_{qg}\otimes\PAP^{(0)}_{gq}\right](z) &=
\Cf \tr \bigg[\frac{4--4 z^3-3 z^2+3 z}{3 z}+2 (z+1) \ln (z)\bigg],
\label{eq:PAP0xPAP0}
\\
\nonumber
\begin{split}
\sum_{x\in [q,g]}
\left[\PAP^{(0)}_{qx}\otimes\PAP^{(0)}_{xg}\right](z) &=
\beta_0 \tr \big[z^2+(1-z)^2\big] + \Ca\tr 
\bigg[
1-\frac{31 z^2}{3}+8 z+\frac{4}{3 z}
\\
&\quad
+2 \left(2 z^2-2 z+1\right) \ln (1-z)+2 (4 z+1) \ln  (z)\bigg] \\
&+\Cf\tr \bigg[-3 z^2 + 5z - 2 +2 \left(2 z^2-2 z+1\right) \log
   (1-z)\\
&\quad-\left(4 z^2-2 z+1\right) \log (z)+\frac{3}{2} 
(1-2z+2z^2)\bigg].
\end{split}
\end{align}

Finally, we find it convenient to introduce a number of generalized
splitting functions and anomalous dimensions. They read 
\begin{align}
\begin{split}
\label{eq:calPp}
&
\calP'_{qq}(z) = \Cf \big[4 \DD1(z)+(1-z)-2 (1+z) \ln (1-z)\big], \vphantom{\bigg]}
\\
&
\calP'_{qg}(z) = 
\tr \big[2 \lp z^2 + (1 - z)^2\rp \ln(1-z) + 2 z (1 - z)\big],\vphantom{\bigg]}
\\
&
\calP'_{gq}(z) = 
\Cf\left[z+2\left(\frac{1+(1-z)^2}{z}\right) \ln (1-z)\right],
\\
&
\calP'_{gg}(z) =
\Ca \left[4 \DD1(z)+4 \left(\frac{1}{z}+z(1-z)-2\right) \ln (1-z)
\right], 
\end{split}
\end{align}
for the splitting functions, and 
\begin{gather}
\gamma'_q = \Cf\lp\frac{13}{2}-\frac{2\pi^2}{3}\rp,
~~~~~~
\gamma'_g = \Ca\lp\frac{67}{9}-\frac{2\pi^2}{3}\rp
-\frac{23}{9}\tr\nf,
\label{eq:gaP}
\\
\gamma_{k_\perp,g} = -\frac{\Ca}{3}+\frac{2}{3}\tr\nf,
\nonumber
\end{gather}
for the anomalous dimensions.
We also use the following quantities
\bes
&\delta_g = \Ca\lp -\frac{131}{72}+\zeta_2\rp
+\frac{23}{36}\tr\nf + \beta_0\ln(2),
\\
&\delta_{k_\perp,g} = \Ca\lp\frac{13}{36}-\frac{\ln(2)}{3}\rp
+\tr\nf\lp-\frac{13}{18}+\frac{2\ln(2)}{3}\rp.
\end{split}
\ee

\section{Partition functions for NNLO calculations}
\label{sec:app_damp}
In this appendix, we report partition functions that we used in our calculations.
They have the same form as those used in 
Refs.~\cite{Caola:2019nzf,Caola:2019pfz}. They read
 \begin{align}
 \begin{split}
&~~w^{51,61} = \frac{\eta_{54}\eta_{64}}{d_5 d_6}
\lp 1 + \frac{\eta_{51}}{d_{5641}} +
\frac{\eta_{61}}{d_{5614}}\rp,
\\
&~~w^{54,64} = \frac{\eta_{51}\eta_{61}}{d_5 d_6}
\lp 1 + \frac{\eta_{54}}{d_{5614}} +
\frac{\eta_{64}}{d_{5641}}\rp,
\end{split}
\\[10pt]
&w^{51,64} = \frac{\eta_{54}\eta_{61}\eta_{56}}{d_5 d_6 d_{5614}}, \quad
w^{54,61} = \frac{\eta_{51}\eta_{64}\eta_{56}}{d_5 d_6 d_{5641}},
\end{align}
where
\be
d_{i=5,6} = \eta_{i1}+\eta_{i4},~~~
d_{5614} = \eta_{56} + \eta_{51} + \eta_{64},~~~
d_{5641} = \eta_{56} + \eta_{54} + \eta_{61}.
\ee
We remind the reader that in our notation
\be
\eta_{ij} = (1-\cos\theta_{ij})/2,
\ee
where $\theta_{ij}$ is the angle between the directions of partons $i$ and $j$.
We also recall that throughout this paper we use the notation
\be
\wt^{5i,6j}_{6||k} = \lim_{\eta_{6k}\to 0} w^{5i,6j}.
\ee

\section{Partitioning-dependent integrals}
\label{sec:app_partint}

In this appendix we comment on the computation of partition-dependent angular integrals
that appear  in the  NNLO subtraction terms.
They read (c.f.  Eq.~\eqref{eq3.44})
\begin{align}
\label{eq:angular_int_def}
\langle \mathcal{O} \rangle_{S_5} \equiv -\epsilon \bigg[\frac{1}{8\pi^2} \frac{(4\pi)^\epsilon}{\Gamma(1-\epsilon)} \bigg]^{-1} 4^\epsilon \int \frac{\textrm{d}\Omega_5^{d-1}}{2(2\pi)^{d-1}} \, \frac{\rho_{41}}{\rho_{51} \rho_{54}} \ \mathcal{O} \, ,
\end{align}
where the function ${\cal O}$ has a residual dependence on the partitioning. 

As an example of computations required in such cases, 
we consider the angular integral that appears in the sum of triple-collinear
sectors $w^{51,61} \theta^{(a,c)}$ and double-collinear sectors $w^{54,61}$.
As we explained in Section~\ref{sec:ns_der}, we need $\langle \Delta_{61} \rangle_{S_5}$, with
\begin{align}
\label{eq:angular_int_exampl}
\Delta_{61} = \widetilde{w}^{54,61}_{6\parallel 1} + \left(\frac{\rho_{51}}{4}\right)^{-\epsilon} \widetilde{w}^{51,61}_{6\parallel 1} \,.
\end{align}
There we have shown  that the dependence of $\langle \Delta_{61} \rangle_{S_5}$ on the partitioning 
starts at $\mathcal{O}(\epsilon^2)$; this is
tantamount to the independence of $1/\epsilon$ poles  in the 
double-real contribution
to the physical cross section on the partitioning.
Below we explain  how $\langle \Delta_{61} \rangle_{S_5}$ can be calculated. 

To compute the integral, we follow  the discussion in Section~\ref{sec:ns_der} and write
\begin{align}
\label{eq:angular_intergal_reg}
\langle \Delta_{61} \rangle_{S_5} = \langle (C_{51} + C_{54}) \Delta_{61} \rangle_{S_5} + \langle (I - C_{51} - C_{54}) \Delta_{61} \rangle_{S_5} \, .
\end{align}
The first term reads 
\begin{align}
\label{eq:angular_int_singular}
\langle (C_{51} + C_{54}) \Delta_{61} \rangle_{S_5} = -\epsilon \bigg[\frac{1}{8\pi^2} \frac{(4\pi)^\epsilon}{\Gamma(1-\epsilon)} \bigg]^{-1} 4^\epsilon \int \frac{\textrm{d}\Omega_5^{d-1}}{2(2\pi)^{d-1}} \,  \left[ \frac{1}{\rho_{54}} + \frac{1}{\rho_{51}} \left(\frac{\rho_{51}}{4}\right)^{-\epsilon} \right] \, .
\end{align}
The first  integral  in Eq.~(\ref{eq:angular_int_singular}) is computed using  Eq.~\eqref{eq2.26}; the second one evaluates to  
\begin{align}
\label{eq:angular_int_1dir_ep}
\int \frac{\textrm{d}\Omega_5^{d-1}}{2(2\pi)^{d-1}} \, \frac{1}{\rho_{51}} \left(\frac{\rho_{51}}{4}\right)^{-\epsilon} = - \frac{1}{\epsilon} \bigg[\frac{1}{8\pi^2} \frac{(4\pi)^\epsilon}{\Gamma(1-\epsilon)}\bigg]\bigg[\frac{2^\epsilon}{2}\frac{\Gamma(1-\epsilon)\Gamma(1-2\epsilon)}{\Gamma(1-3\epsilon)}\bigg] 2^{-2\epsilon} \, .
\end{align}
The second term on the right-hand side of Eq.~\eqref{eq:angular_intergal_reg} is fully regulated and can be expanded in $\epsilon$.
Using the ``completeness'' relation  for the partition functions Eq.~\eqref{eq:completeness} we write it as 
\begin{align}
\label{eq:angular_int_example_reg_part}
\begin{split}
&\langle (I - C_{51} - C_{54}) \Delta_{61} \rangle_{S_5} \\
&= - \epsilon  \bigg[\frac{1}{8\pi^2} \frac{(4\pi)^\epsilon}{\Gamma(1-\epsilon)} \bigg]^{-1} 4^\epsilon \int \frac{\textrm{d}\Omega_5^{d-1}}{2(2\pi)^{d-1}} \\
&\quad\times \bigg[ \left( \frac{\rho_{41}}{\rho_{51}\rho_{54}} - \frac{1}{\rho_{51}} - \frac{1}{\rho_{54}} \right) +  \frac{1}{\rho_{51}}  \left( \left(\frac{\rho_{51}}{4}\right)^{-\epsilon} - 1 \right) \left( \frac{\rho_{41}}{\rho_{54}} \widetilde{w}^{51,61}_{6\parallel 1} - 1 \right) \bigg] \, .
\end{split}
\end{align}
Note that the two  terms in brackets in the integrand in Eq.~\eqref{eq:angular_int_example_reg_part} are independently finite. 
The first term can be
computed using known integrals  Eqs.~(\ref{eq2.26},~\ref{eq:angular_int_1dir_ep}).
The second term  is the only one that  depends on the chosen partitioning.

To proceed further, we expand the second term in the integrand  in  Eq.~\eqref{eq:angular_int_example_reg_part} and obtain
\begin{align}
\label{eq:angular_int_expanded}
\begin{split}
&- \epsilon  \bigg[\frac{1}{8\pi^2} \frac{(4\pi)^\epsilon}{\Gamma(1-\epsilon)} \bigg]^{-1} 4^\epsilon \int \frac{\textrm{d}\Omega_5^{d-1}}{2(2\pi)^{d-1}} \, \frac{1}{\rho_{51}}  \left( \left(\frac{\rho_{51}}{4}\right)^{-\epsilon} - 1 \right) \left( \frac{\rho_{41}}{\rho_{54}} \widetilde{w}^{51,61}_{6\parallel 1} - 1 \right) \\
&= - \epsilon^2 \times \frac{1}{2\pi} \int \textrm{d}^3 \Omega_5 \ \left( \frac{1}{\rho_{51}} \left[ \frac{\rho_{41}}{\rho_{51} + \rho_{54}} - 1 \right] + \frac{\rho_{41} }{(\rho_{51} + \rho_{54})^2}\right) \ln\left(\frac{\rho_{51}}{4}\right) + \mathcal{O}(\epsilon^3) \, ,
\end{split}
\end{align}
where we have used the explicit form of the partition function
\begin{align}
\widetilde{w}^{51,61}_{6\parallel 1} = \frac{\rho_{54}}{\rho_{51} + \rho_{54}} \left(1+\frac{\rho_{51}}{\rho_{51} + \rho_{54}} \right) \, ,
\end{align}
see Appendix~\ref{sec:app_damp}.

To compute the remaining integral, it is convenient to choose the $z$-axis along the direction of the
vector $\vec n_1$ since, with this choice, $\log \rho_{51}$ becomes independent of the azimuthal angle $\varphi_5$.
Remaining integrals over $\varphi_5$ can be performed using the well-known formulas
\begin{align}
\label{eq:phi_intergals}
\int \textrm d \varphi_5 \frac{1}{(a - b \cos\varphi_5)^n} = 2 \pi \times \left\{ 
\begin{array}{ll}
1, &  \  n = 0 \\
(a^2 - b^2)^{-\frac{1}{2}}, &  \ n = 1 \\
a (a^2 - b^2)^{-\frac{3}{2}},  &  \ n = 2 
\end{array} 
\right. \, .
\end{align}
One can explicitly check that after integration over $\varphi_5$, only squares of 
$\sin\theta_5$ appear; this implies that the remaining integrands contain square roots of polynomials
of $\cos \theta_5$.  These roots can be rationalized and integrated. 
Combining everything and expanding remaining terms in $\epsilon$, 
the result reads
\begin{align}
\langle \Delta_{61} \rangle_{S_5} &= \frac{3}{2} + \epsilon \left( 
\frac{\ln (2)}{2} - 2 \ln (\eta_{14}) \right) + \epsilon^2 
\bigg( - \frac{\pi^2}{3} - \ln (2) + \frac{\ln^2 (2)}{4}  
- \frac{\ln\left(\frac{1+\sqrt{1-\eta_{14}}}{1-\sqrt{1-\eta_{14}}}\right)}
{2 \sqrt{1-\eta_{14}}} \nonumber\\
&\quad + \frac{\ln(\eta_{14})}{2} - \ln (2) \ln(\eta_{14})
 + \frac{3\ln^2(\eta_{14})}{2}
+\frac{5}{2} \textrm{Li}_2(1-\eta_{14}) \bigg)  + \mathcal{O}(\epsilon^3) \, .
\label{eq:angular_int_example_expanded}
\end{align}

Other integrals that depend on the partition functions and appear in
the subtraction terms can be calculated along the same lines. 
For final-state partitions, we need $\la \Delta_{64}\ra_{S_5}$ with 
\be
\Delta_{64} = \wt^{51,64}_{6||4} + \lp\frac{\rho_{54}}{4}\rp^{-\ep}
\wt^{54,64}_{6||4}.
\ee
Thanks to the symmetry of the damping factors, it is immediate to see
that $\la \Delta_{64}\ra_{S_5} = \la \Delta_{61}\ra_{S_5}$.
In 
sectors $b$ and $d$, we also require $\la \Delta_{65}\ra_{S_5}$ with
\be
\Delta_{65} = \sum_{i\in[1,4]}\wt^{5i,6i}_{6||5} 
\lp\frac{\eta_{5i}}{1-\eta_{5i}}\rp^{-\ep}.
\ee
Using manipulations similar to the ones just described, we obtain
\be
\la \Delta_{65}\ra_{S_5} = 1 - 2\ep \ln(\eta_{14}) + 
\ep^2 \bigg(\Li_2\big[(1-\eta_{14})^2\big] + 2\ln^2(\eta_{14}) 
-\frac{2}{2-\eta_{14}}\bigg) + \mathcal O(\ep^3).
\label{eq:c12}
\ee

In our final formulas, we denote the $\mathcal O(\ep^2)$ part of
$\la \Delta_{65}\ra_{S_5}$ and $\la \Delta_{61}\ra_{S_5}$
as $1/2\la\Delta_{61}\ra_{S_5}''$ and $1/2\la\Delta_{65}\ra_{S_5}''$, respectively.
They can easily be read off Eqs.~(\ref{eq:angular_int_example_expanded},\ref{eq:c12}), giving 
\bes
\la\Delta_{61}\ra_{S_5}'' = \la\Delta_{64}\ra_{S_5}'' &= 
- \frac{2\pi^2}{3} - 2\ln (2) + 
\frac{\ln^2 (2)}{2}  
- \frac{\ln\left(\frac{1+\sqrt{1-\eta_{14}}}{1-\sqrt{1-\eta_{14}}}\right)}
{\sqrt{1-\eta_{14}}}
 + \ln(\eta_{14}) 
\\
&- 2\ln (2) \ln(\eta_{14})
 + 3\ln^2(\eta_{14})
+5\Li_2(1-\eta_{14}),
\\
\la\Delta_{65}\ra_{S_5}'' &= 
2\Li_2\big[(1-\eta_{14})^2\big] + 4\ln^2(\eta_{14}) 
-\frac{4}{2-\eta_{14}}.
\end{split}
\label{eq:D2def}
\ee

Finally, we also require the following finite integral
\be
\la r^\mu r^\nu \ra_{\rho_5} \equiv
\sum_{i\in[1,4]}
\int\frac{\d^3\Omega_5}{2\pi}
\left[\lp \frac{n_1\cdot r^{(i)}}{n_1\cdot n_5}
-\frac{n_4\cdot r^{(i)}}{n_4\cdot n_5}\rp^2
-2\frac{n_1\cdot n_4}{(n_1\cdot n_5)(n_4\cdot n_5)}
\right]\wt^{5i,6i}_{6||5},
\ee
where $n_i = p_i/E_i$ and the $r^{(i)}$ vector has been introduced in the
main text (see the discussion around Eq.~\eqref{eq:rdef}). Using the 
explicit formula for the partition functions shown in Appendix~\ref{sec:app_damp}, we obtain
\be
\la r^\mu r^\nu \ra_{\rho_5} = 2 \left[ \frac{1}{2-\eta_{14}} -1 
-\ln(2-\eta_{14})\right].
\label{eq:spint}
\ee

\end{document}